\documentclass[journal,comsoc]{IEEEtran}

\usepackage[normalem]{ulem}
\usepackage{amssymb,color}
\usepackage{amsmath}
\usepackage{bbm}
\usepackage{graphicx}
\usepackage{epstopdf}
\usepackage{amsthm,amsfonts}
\usepackage{tikz}

\usepackage{subcaption}

\usepackage{hhline}
\usepackage{multirow}

\usepackage{accents}

\usepackage{algorithm}
\usepackage{algorithmic}
\usepackage{psfrag}
\usepackage{cite}
\usepackage{pifont}
\usepackage[font={footnotesize}]{caption}

\allowdisplaybreaks

\makeatletter
\newcommand{\oset}[3][0ex]{%
  \mathrel{\mathop{#3}\limits^{
    \vbox to#1{\kern-2\ex@
    \hbox{$\scriptstyle#2$}\vss}}}}
\makeatother

\begin{document}
\title{Impact of Network Geometry on Large Networks with Intelligent Reflecting Surfaces}

\author{
\IEEEauthorblockN{\large  Konpal Shaukat Ali$^*$, \textit{Member, IEEE}, Martin Haenggi$^{\dagger}$, \textit{Fellow, IEEE}, Arafat Al-Dweik$^{\ddagger}$, \textit{Senior Member, IEEE} and Marwa Chafii$^*$, \textit{Senior Member, IEEE}.}

\thanks{$^{*}$The authors are with the Engineering Division, New York University (NYU) Abu Dhabi, 129188, UAE (Email: \{konpal.ali, marwa.chafii\}@nyu.edu). M. Chafii is also with NYU WIRELESS, NYU Tandon School of Engineering, Brooklyn, 11201, NY.

$^{\dagger}$The author is with the Department of Electrical Engineering, University of Notre Dame, USA. (Email: mhaenggi@nd.edu).

$^{\ddagger}$The author is with the Department of Electrical and Computer Engineering, Khalifa University, Abu Dhabi, UAE (Email: arafat.dweik@ku.ac.ae) and is also with the Department of Electrical and Computer Engineering, Western University, London, ON, Canada. (Email: dweik@fulbrightmail.org).

 }}

\maketitle

\begin{abstract}
In wireless networks assisted by intelligent reflecting surfaces (IRSs), jointly modeling the signal received over the direct and indirect (reflected) paths is a difficult problem. In this work, we show that the network geometry (locations of serving base station, IRS, and user) can be captured using the so-called triangle parameter $\Delta$. We introduce a decomposition of the effect of the combined link into a signal amplification factor and an effective {channel power} coefficient $G$. The amplification factor is monotonically increasing with both the number of IRS elements $N$ and $\Delta$. For $G$, since an exact characterization of the distribution seems unfeasible, we propose three approximations depending on the value of the product $N\Delta$ {for Nakagami fading and the special case of Rayleigh fading}. For two relevant models of IRS placement, we prove that their performance is identical if $\Delta$ is the same given an $N$. {We also show} that no gains are achieved from IRS deployment if $N$ and $\Delta$ are both small. We further compute bounds on the diversity gain to quantify the channel hardening effect of IRSs. Hence only with a judicious selection of IRS placement and other network parameters, non-trivial gains can be obtained.

\end{abstract}


\section{Introduction}
\subsection{Background}

With the shift towards smart environments that involve a plethora of interconnected devices, new efficient technologies are required to meet the growing network demands. Previously, the focus had been on efficient spectrum reuse techniques, massive multiple input multiple output (MIMO), millimeter wave communication and heterogeneous network densification. However, they do not allow control over the rapidly changing wireless propagation environment. With the new challenges and networking trends of the future, radically new communication paradigms are required in the physical layer \cite{6Gvision,IRS6}. In this context, {intelligent reflecting surfaces (IRSs)} are gaining attention as a means to attain control over the propagation environment \cite{IRSmag1,IRSmag2}.



An IRS is a surface comprised of a large number of nearly passive reflecting elements that have reconfigurable parameters. {It is anticipated to be low-cost \cite{IRSmag1,IRSmag1_9}.} A remote smart controller adapts the function of each of the IRS elements. Typically, each element of an IRS can manipulate the {incoming} signal by reflection in a desired direction, {phase shifts,} absorption or polarization {in addition to allowing the signal} to pass through. This way, the function of each element can be adapted according to the wireless propagation environment to achieve different goals such as signal enhancement, interference cancellation or jamming. {While a base station (BS) and {user equipment} (UE) typically only share a direct {BS-UE} link, an IRS is a means of adding an indirect {BS-IRS-UE} link.} Note that in contrast to massive MIMO and network densification, an IRS does not generate additional signals but instead optimizes the reflection of existing signals, thereby not incurring additional power consumption \cite{IRS2} {other than that required by the remote smart controller}. As the IRS focuses the signal in a particular direction, it does not create additional network interference of significance\footnote{{For other receivers, the IRS acts like a standard (non-intelligent) reflecting surface whose effects {are usually} incorporated in the fading model.}}. {In our work, we focus on using the IRS for {phase adjustment that results in constructive superposition at the receiver} to enhance the desired signal.}


\subsection{Related Work}

Recently, there has been a lot of interest in studying IRSs in different contexts. {A number of works have focused on phase estimation of the IRS to achieve different goals \cite{IRSsurvey_5,IRSsurvey_51}.} Works such as \cite{IRSphase1} have focused on a practical phase shift model. The work in \cite{IRS7} focused on imperfect phase estimation in setups deploying IRSs and showed that accurate knowledge of phase shifts is not necessary to closely approach the ideal performance. {The impact of imperfect phase estimation on the bit error rate (BER) and outage probabilities was considered in \cite{Arafat_BER_IRS}. It was shown that increasing the number of IRS elements may degrade the BER in certain scenarios. In \cite{sobia_IRS} it was shown in that the number of activated reflectors should be optimized to minimize the BER in the presence of imperfect phase control.} The capacity of IRS-assisted networks with imperfect phase estimation and compensation error was studied in \cite{Arafat_capacity_IRS}. IRSs have also been used in conjunction with other technologies to reap benefits from the physical layer. {In \cite{IRS6} it was shown that IRS-based index modulation can result in high data rates and low error rates.} In \cite{IRS_NOMA1} IRS-assisted non-orthogonal multiple access (NOMA) transmissions were studied. In \cite{IRS2} it was shown that by deploying IRSs in MISO systems, massive MIMO like gains can be achieved using fewer antennas, thereby reducing power consumption. Performance analysis in the presence of a direct link was done in \cite{arafatIRS_1,arafatIRS_2}.

Most of the works on IRSs have focused on a setup with a single or few cells. {For NOMA,} it has been shown that studying such setups can lead to inaccurate deductions as well as result in suboptimum parameter selection that can hurt performance in a real network \cite{my_nomaMag}. With networks becoming more dense and {increasingly interference-limited}, studying a large network model that accurately mimics a real world network is of great value. In this context, stochastic geometry provides a unified mathematical paradigm for modeling large wireless cellular networks and characterizing their operation while taking intercell interference into account \cite{MH_Book2,mh_Asymptotics,myNOMA_tcom}. Works such as \cite{IRS_SG5,IRS_SG6,IRS_SG2,IRS_SG7,IRS_SG8} have studied IRSs in the context of large networks. {In \cite{IRS_SG5}, a Poisson bipolar network is used to model the BSs and UEs. IRSs form an independent Poisson point process (PPP) and are represented using line segments with random lengths and orientations. A UE connects to the BS either via the direct link or over the indirect link, but never both. Rayleigh fading is assumed over both the direct (BS-UE) and the entire {collated} indirect link (BS-IRS-UE); {since the indirect link is comprised of the two {sub-links} BS-IRS and IRS-UE, this} not accurate.} {In \cite{IRS_SG6}, the received signal is from either the direct link or the indirect link, which operate on different frequencies with different transmission powers. The IRS is assumed to be located in the middle of the BS and UE; {it is, however, unclear how the signal could be reflected towards a point lying exactly behind the IRS}.} In \cite{IRS_SG2}, the distributions of the signal and interference components are derived separately and used to characterize the statistics of the signal-to-interference-and-noise ratio (SINR). Additionally, the impact of the signals received from both the direct and indirect links is incorporated into the analysis. It is shown that in their model, IRS deployment significantly boosts the signal power but only causes a marginal increase in the network interference. {In \cite{IRS_SG7} the locations of the transmitters and IRSs are modeled using a Gauss-Poisson process. The impact of the signal from both the direct and indirect link is considered, and the signal power received from both links is approximated using a gamma random variable via moment matching.} {In \cite{IRS_SG8}, a MIMO network is considered where IRSs are modeled using a Mat\'ern cluster process around the BS. All of the IRSs in the cluster aid transmission and their elements are shared by all of the UEs. The signal from both the direct and indirect link is considered and the statistics of each signal component are analyzed independently for the SIR statistics.} 

\subsection{Proposed Approach}
{IRS-aided setups are difficult to analyze because of the presence of the signals from both the direct and the indirect link. Our {approach} is to tackle this problem by providing a comprehensive analytical framework and proposing models for the joint channel observed at the receiver as a result of the direct and indirect link. The proposed models for the joint channel are used to obtain accurate approximate {expressions} for the CDF of the signal-to-interference ratio (SIR). {Different from previous works,} we bring the SIR into a general form where the combined {channel for the signal} is in the form of a path loss factor multiplied by a signal amplification term and a unit-mean \emph{effective channel power coefficient {(ECPC)}} $G$ {that incorporates the impact of both the direct and indirect link}. Modeling the channels of the reflected paths from the indirect link jointly with that from the direct link {is insightful and simplifies the analytical formulation, but} calculating the exact statistics of the ECPC $G$ is difficult. They depend on the model used for IRS {placement} and, consequently, the triangle parameter $\Delta$ that captures the geometry of the triangle formed by the BS, IRS and UE, the channels on the direct and indirect links, as well as the number of IRS elements $N$. {We consider Nakagami fading on the direct and indirect links and single out some results specific to the special case of Rayleigh fading. With Nakagami fading, to approximate the statistics of $G$ that allow tractability in computing the SIR statistics, we propose using unit-mean Erlang RVs: 1) $G_{\rm Erl}^{\rm s, Nak}$ when $N\Delta$ is small, 2) $G_{\rm Erl}^{\rm m, Nak}$ for medium values of $N\Delta$ and 3) $G_{\rm Erl}^{\rm l, Nak}$ when $N\Delta$ is large, each with different parameters. With Rayleigh fading, this becomes a unit-mean: 1) exponential random variable (RV) $G_{\rm exp}^{\rm s, Rayl}$ when $N\Delta$ is small, 2) Erlang RV $G_{\rm Erl}^{\rm m, Rayl}$ for medium values of $N\Delta$ and 3) Erlang RV $G_{\rm Erl}^{\rm l, Rayl}$ when $N\Delta$ is large.} We propose and study two models for IRS placement to analyze the impact that the distances in the indirect link have on the performance. {To the best of our knowledge, a joint channel model comprising the direct and indirect link has not been {proposed}.} The derived SIR analysis is used to study the throughput and calculate the diversity gain. {The scenario without IRS is used to benchmark the gains obtained.}

\subsection{Contributions}
{The contributions of this work are summarized as follows:
\begin{itemize}

\item {We show that IRSs have a two-fold impact on the channel: they result in signal amplification and they improve the fading conditions, which is captured by the ECPC $G$.} Significant gains can be reaped from a network that deploys IRSs compared to one that does not. 
\item {We show that the network geometry, based on the locations of the BS, IRS and UE, can be captured by the {so-called} triangle parameter $\Delta$.} The impact of $\Delta$ on the performance reflects the significance of the impact of IRS placement, network density and relative path loss. 
\item {We show that when $N\Delta$ is small, IRS-aid has a negligible impact on fading improvement and the statistics of $G$ can be approximated as they would in the case without IRS. When $N\Delta$ is larger, we propose approximations for the CDF of $G$.}%
\item {While our analysis and proposed statistics for $G$ hold for any IRS placement,} we propose and analytically study two models for IRS placement. {In Model I, the IRS is placed at a fixed distance from the UE in a random direction and we derive the distribution of the link distance between the BS and IRS.} In Model II, the IRS is placed equidistant from the BS and the UE.
\item {We show that the proposed approximate statistics of $G$ only depend on the value of $N\Delta$. Further, for a given $N$, irrespective of the model, the {throughput and reliability} performance is identical for the same $\Delta$.}
\item To measure the improvement in fading, we calculate bounds on the diversity gain using the statistics of each approximation of $G$. {In the case of Rayleigh fading {($\mu=1$), two lower bounds on the diversity are derived as well as an upper bound, which holds for all $N\Delta$}. In the case of Nakagami fading {with $\mu>1$}, three upper bounds on the diversity are derived.} %
\item We show that the signal amplification factor in the SIR is $\Theta(N^2)$ as $N \to \infty$, {and thus} the throughput increases linearly with $N$. This is in line with the findings in \cite{IRS0}.
\end{itemize}
}

\textit{Notation:} {We denote vectors using bold text. The Euclidean norm of the vector $\textbf{z}$ is denoted by $\|\textbf{z}\|$. The ball centered at $\textbf{z}$ with radius $R$ is denoted by $b(\textbf{z},R)$.} The ordinary hypergeometric function is denoted by ${}_2F_1$. The CDF (CCDF, PDF) of the RV $X$ is denoted by $F_X$ ($\bar{F}_X$, $f_X$). {The Laplace transform (LT) of the PDF of the RV $X$ is denoted by $\mathcal{L}_X(s)=\mathbb{E}[e^{-sX}]$. The Pochhammer polynomial is $(a)_n= \prod_{i=0}^{n-1} (a+i)$.}

\section{{General System Model and SIR Expression}}\label{SysMod}

\subsection{{Network Model}}
\begin{figure}[thb]
\begin{minipage}[thb]{\linewidth}
\centering\includegraphics[width=0.6\linewidth]{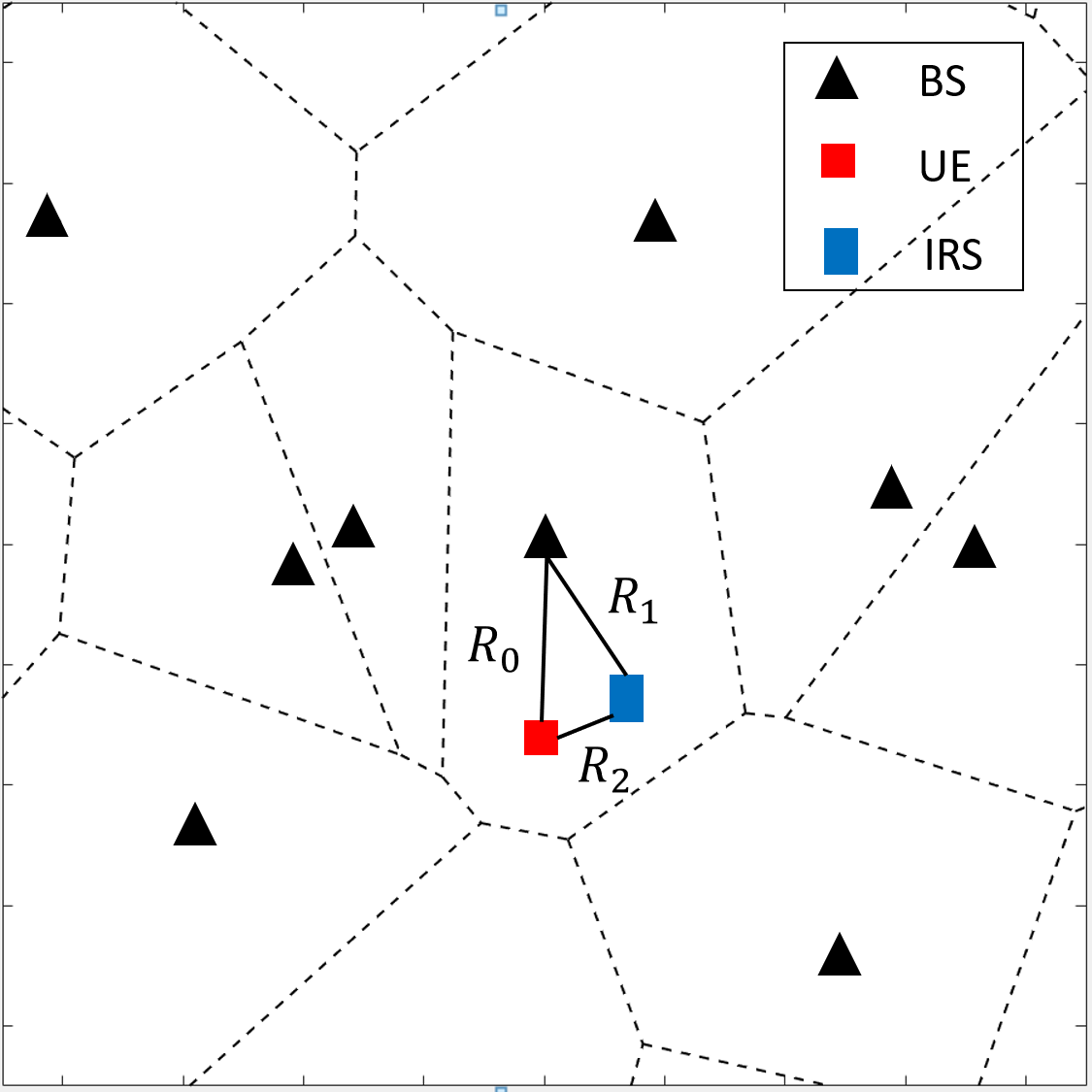}
\end{minipage}
\caption{A realization of the network. {The tBS, tUE and tIRS are shown in the typical cell.}}\label{genIRS}
\end{figure}
We consider a downlink cellular network where {isotropic BSs} are distributed according to an arbitrary {stationary} point process $\Phi \subset \mathbb{R}^2$ and UEs are distributed uniformly at random in the cells. A BS assigns each UE a unique time-frequency resource block to avoid interference between the messages to the UEs in a cell. We focus on one such resource block. A BS transmits to its UE with power $P$; without loss of generality, we assume $P=1$. {As usual we study the performance of the typical UE, referred to as the tUE, being served by the typical BS, referred to as the tBS, in the typical cell.} {The tBS and tUE will be specified in the next section.} {As cellular networks are interference-limited, the focus of this work is on the statistics of the SIR instead of the SINR.} 



 The IRSs form a stationary point process that may depend on $\Phi$. Each IRS is assumed to have $N$ elements. The IRS assisting the tUE is referred to as the typical IRS and denoted by tIRS. {We assume that the reflection from the tIRS is directed towards the tUE.} Additionally, each IRS operates only on the resource block of the UE it is assisting. {In this work, we assume that only one IRS assists a UE in one resource block. Thus, the tIRS is the only IRS assisting the tUE.} Note that in this work the signals of interest received at the tUE from both the {\emph{direct link}, tBS-tUE, and the \emph{indirect link}, tBS-tIRS-tUE,} is taken into consideration.




{We consider i.i.d. Nakagami fading with shape parameter $\mu$ and spread parameter $\omega$ on each of the direct link and sub-links of the indirect link, referred to as the D\&I links from hereon. {The corresponding channel power coefficient is gamma distributed with parameters $k=\mu$ and $t=\omega/\mu$, with mean $kt$ and variance $kt^2$.} When $\mu=1$, the fading becomes Rayleigh. We assume that the fading coefficient between any two nodes results in a unit-mean channel power coefficient. {For Nakagami fading, this implies that the $\omega=1$ so that corresponding channel power coefficient has a unit-mean gamma distribution with parameters $\mu$ and $1/\mu$.} All of the interfering links experience i.i.d. Rayleigh fading and the corresponding channel power coefficient has a unit-mean exponential distribution.} A power-law path loss model is considered where the signal decays at the rate {$\left(\frac{r}{\ell_0}\right)^{-\eta}$} with distance $r$, {where $\ell_0$ is the reference distance}, $\eta>2$ denotes the path loss exponent and $\delta=\frac{2}{\eta}$. {Fixed-rate transmissions are used such that a message is sent with {bandwidth-normalized} transmission rate $\log(1+\theta)$ corresponding to an SIR threshold of $\theta$. Such transmissions result in a throughput that is lower than the transmission rate because of possible outages.}


In our work, IRSs are used to optimize the phase of the reflected signal from the BS. We assume the $i^{\rm th}$ element of the IRS, where $1\leq i \leq N$, multiplies the incoming signal with $e^{-j \alpha_i}$. {The phase coefficients $\alpha_i$ of the IRS elements are {chosen} such that the signal from the indirect link and the signal from the direct link add constructively at the receiver; {which is the optimum}.} 


The distance between the tBS and the tUE is denoted by $R_0$, the distance between the tBS and the tIRS is denoted by $R_1$, and the distance between the tIRS and tUE is denoted by $R_2$ as shown in Fig. \ref{genIRS}. Let $\textbf{g}_{1}=[g_{1,1} e^{-jz_{1,1}},\ldots,g_{N,1}e^{-jz_{N,1}}]$ denote the $N$-dimensional vector of channel coefficients from the tBS to the $N$ elements of the tIRS. {For $1 \leq i \leq N$, $g_{i,1}$ follows a Nakagami distribution with parameters $\mu$ and $\omega=1$} and $z_{i,1}$ is uniformly distributed on $[0, 2\pi]$. {Similarly, the $N$-dimensional vector of channel coefficients from the tIRS to the tUE is denoted by $\textbf{g}_{2}=[g_{1,2} e^{-jz_{1,2}},\ldots,g_{N,2}e^{-jz_{N,2}}]$. Again, for $1 \leq i \leq N$, $g_{i,2}$ follows a {Nakagami distribution with parameters $\mu$ and $\omega=1$} and $z_{i,2}$ is uniformly distributed on $[0, 2\pi]$.} {We assume that the elements of the IRS are sufficiently spaced apart (by at least half the
wavelength) and the environment is richly scattered so that the fading between the $N$ elements of the IRS is independent \cite{IRS7,Emil_IRSvsDF}.} Let $\textbf{g}_{0}=g_0 e^{-jz_0}$ denote the direct channel between the tBS and tUE; $g_0$ follows a {Nakagami distribution with parameters $\mu$ and $\omega=1$} and $z_0$ is uniformly distributed on $[0, 2\pi]$. Since the tIRS adjusts the signal from the indirect link, we represent the impact of the adjustment from the $N$ elements of the IRS using an $N \times N$ matrix $\textbf{W}=\text{diag}( e^{-j \alpha_1}, \ldots,  e^{-j \alpha_N} )$, where $\alpha_i=z_0-z_{i,1}-z_{i,2}$. 


\subsection{{SIR Expression}}
The SIR at the tUE {for any IRS placement model} can be written as
\begin{align}
&{\rm SIR}= \frac{ \left|\left(\!\frac{R_0}{\ell_0}\!\right)^{-\frac{\eta}{2}}  \textbf{g}_0 + \left(\!\frac{R_1}{\ell_0}\!\right)^{-\frac{\eta}{2}} \left(\!\frac{R_2}{\ell_0} \!\right)^{-\frac{\eta}{2}} \textbf{g}_1^T \textbf{W} \textbf{g}_2 \right|^2}{ I } \nonumber \\
&\!= \frac{ \left| \! \left(\!\frac{R_0}{\ell_0}\!\right)^{\!-\frac{\eta}{2}} \!g_0 e^{-jz_0} \! + \! \left(\!\frac{R_1 \!R_2}{\ell_0^2} \!\right)^{\!-\frac{\eta}{2}} \! \sum\limits_{i=1}^N \! g_{i,1}   g_{i,2}  e^{-j (z_{i,1} + z_{i,2}+ \alpha_i)}  \! \right|^2}{ I } \! \nonumber \\
&\stackrel{(a)}= {\frac{ \left(\!\frac{R_0}{\ell_0} \!\right)^{-\eta} \left| g_0  + \left(\! \frac{R_0 \ell_0}{R_1 R_2} \!\right)^{\frac{\eta}{2}} \! \sum\limits_{i=1}^N g_{i,1} g_{i,2}     \right|^2}{ I   },} \; \label{sinr_gen}
\end{align}
{where (a) follows from $\alpha_i=z_0-z_{i,1}-z_{i,2}$, i.e., the signals received from the direct and indirect link are in phase, $|e^{-jz_0}|=1$ and rewriting the equation.} The intercell interference from the other BSs in the network, {referred to as simply the interference,} is $I=\sum_{\textbf{x} \in \Phi}   \left(\frac{\|\textbf{y}\|}{\ell_0}\right)^{-\eta} h_{\textbf{y}} $, where $\textbf{y}=\textbf{x}-\textbf{u}$ and $\textbf{u}$ is the location of the tUE. The channel power coefficient between the interfering BS at \textbf{x} and tUE is $h_{\textbf{y}}\sim \exp(1)$.



While the network has IRSs other than the tIRS operating on the same resource block as the tUE, we do not need to explicitly account for the interference from these IRSs in our SIR expression. {This is because Rayleigh fading already assumes many reflected signals to combine non-coherently at the receiver. An ``unmatched'' IRS simply behaves as one such reflector for the signal from interfering BSs and is incorporated in $I$.} {Similarly, we do not {need to separately account for} the interference signal that originates from interfering BSs and is multiplied with \textbf{W} by the tIRS before being received at the tUE. This is because the tIRS is not matched to the interfering BSs and thus acts as another reflector for the signals from the interfering BSs; these reflections too are incorporated in $I$ by the Rayleigh fading assumption.} {Works such as \cite{IRS_SG8} also emphasized that such signals play a negligible role.}

{Next we formally define the important triangle parameter.}

{\textbf{\emph{Definition 1:} Triangle parameter -} The triangle parameter, denoted by $\Delta$, captures the {local} geometry of the triangle formed by the {tBS, tIRS and tUE} and is defined as
{ \begin{align}
\Delta =\left(\frac{R_0 \ell_0}{R_1 R_2}\right)^{\eta}.
\end{align}}
{In the no-IRS case, $R_1=\infty, R_2=\infty$ and thus $\Delta=0$.}

{It should be noted that $\Delta$ captures the ratio of the path gain of the direct link to that of the indirect links. }

{\textbf{\emph{Remark 1:}} Typicality of the tBS-tIRS-tUE triangle means that all the BS-IRS-UE triangles formed in the network have the same statistics as $\Delta$. This implies that although $\Delta$ is defined using the local geometry only, it depends on the global network parameters such as the BS density and is representative for the entire network. }

{Also, we} define the RV $\tilde{G}$ as
{ \begin{align}
\tilde{G}=\Big| g_0 +\Delta^{\frac{1}{2}} \sum_{i=1}^N g_{i,1} g_{i,2}\Big|^2. \label{Y}
\end{align}  }

Then the SIR can be written as
{\begin{align}
{\rm SIR}= \frac{R_0^{-\eta} \ell_0^{\eta} \tilde{G}}{I}.
\end{align}}

{
\textbf{\emph{Lemma 1:}} The mean of $\tilde{G}$ {conditioned on $\Delta$} is {
{ \begin{equation}
\mathbb{E}[\tilde{G} \mid \Delta]= 1 + N \left( 2  \sqrt{\Delta}  \left( \mu_{\rm fad} \right)^3  +  \Delta \left( 1 - \mu_{\rm fad}^4 \right) \right)  +  N^2 \Delta \mu_{\rm fad}^4, \label{E_Y}
\end{equation} }
where $\mu_{\rm fad}$ is the first moment of the fading coefficient; {with Nakagami fading on the D\&I links, $\mu_{\rm fad}=\frac{\Gamma(\mu+1/2)}{\Gamma(\mu)} \sqrt{\frac{1}{\mu}}$.} \\
\textbf{\emph{Proof:}} See Appendix \ref{L1proof}.


{In the case of Rayleigh fading {($\mu=1$)} on the D\&I links, the result in Lemma 1 can be approximated as $\mathbb{E}[\tilde{G} \mid \Delta] \approx 1 + 1.39 N \sqrt{\Delta} + 0.38 N\Delta + 0.62 N^2 \Delta$. This shows that all the coefficients have similar magnitudes and thus all terms are significant.} 
 

%

$\tilde{G}$ captures the impact of fading as well as path loss from both the direct and indirect links. Approximating its statistics is difficult due to its dependence on $N$ and $\Delta$. To facilitate the analysis {and separate the effects of signal amplification and effective fading,} we introduce a scaled version of $\tilde{G}$ that has a unit mean.

\textbf{\emph{Definition 2:} Effective Channel Power Coefficient (ECPC) -} The ECPC, {denoted as $G$,} is defined as the normalized version of $\tilde{G}$ that is unit-mean given $\Delta$, i.e.,
{ \begin{align}
G= \frac{\tilde{G}}{\mathbb{E}[\tilde{G} \mid \Delta]}=\left| \frac{g_0}{\mathbb{E}[\tilde{G} \mid \Delta]^\frac{1}{2}} +{\left(\frac{\Delta}{\mathbb{E}[\tilde{G} \mid \Delta]}\right)}^{\frac{1}{2}} \sum_{i=1}^N g_{i,1} g_{i,2}\right|^2. \label{G_def}
\end{align} } 
{By this definition, $G$ also has unconditional unit mean.} {The ECPC $G$ is thus a unit-mean measure of fading.} The SIR can then be expressed as
{ \begin{align}
{\rm SIR}= \frac{R_0^{-\eta} \ell_0^{\eta} \mathbb{E}[\tilde{G} \mid \Delta] G}{I }. \label{sinr_gen0}
\end{align} }

{
\textbf{\emph{Remark 2:}} The signal component in \eqref{sinr_gen0} is in the form of the path loss term $R_0^{-\eta}\ell_0^{\eta}$ scaled by a fading term which includes $G$, the unit-mean ECPC, and a signal amplification term $\mathbb{E}[\tilde{G}\mid \Delta] \geq 1$.}

The term $\mathbb{E}[\tilde{G}\mid \Delta]=1$ in the case with no IRS (i.e., $N=0$) and increases with $N$ and $\Delta$. The presence of this term in the numerator of the SIR expression in \eqref{sinr_gen0} highlights the direct signal strengthening role of $\mathbb{E}[\tilde{G}\mid \Delta]$. {From Lemma 1, $\mathbb{E}[\tilde{G} \mid \Delta]=\Theta(N^2)$ as $N \to \infty$ and $\mathbb{E}[\tilde{G} \mid \Delta]=\Theta(\Delta)$ as $\Delta \to \infty$. While the impact of increasing $N$ is intuitive, the impact of increasing $\Delta$ on $\mathbb{E}[\tilde{G} \mid \Delta]$ {as well as on the statistics of $G$} implies that the model used (and the corresponding link distances) also plays a vital role in the gains achievable with IRS assistance.} {In particular, the geometry of the triangle formed by the BS, IRS and UE can be adjusted to make $\Delta$ larger and, in turn, further amplify the signal. } 

Using \eqref{sinr_gen0}, the CCDF of the SIR can be written as 
{\small \begin{align}
\mathbb{P}({\rm SIR}>\theta)= \mathbb{P} \left( G > \frac{\theta R_0^{\eta} \ell_0^{-\eta} I}{ \mathbb{E}[\tilde{G}\mid \Delta]}  \right) =\mathbb{E}\left[ \bar{F}_{G} \left( \frac{\theta R_0^{\eta} \ell_0^{-\eta} I}{ \mathbb{E}[\tilde{G}\mid \Delta]}    \right) \right]. \label{ccdfSIR_Y2}
\end{align} }

\section{\textcolor{teal}{SIR Analysis for General Networks}}\label{sec3}
{The analysis in this section is general and applies to any IRS placement model.}

\subsection{On the Distribution of $G$}
\begin{figure}
\begin{minipage}[htb]{\linewidth}
\centering\includegraphics[width=0.95\linewidth]{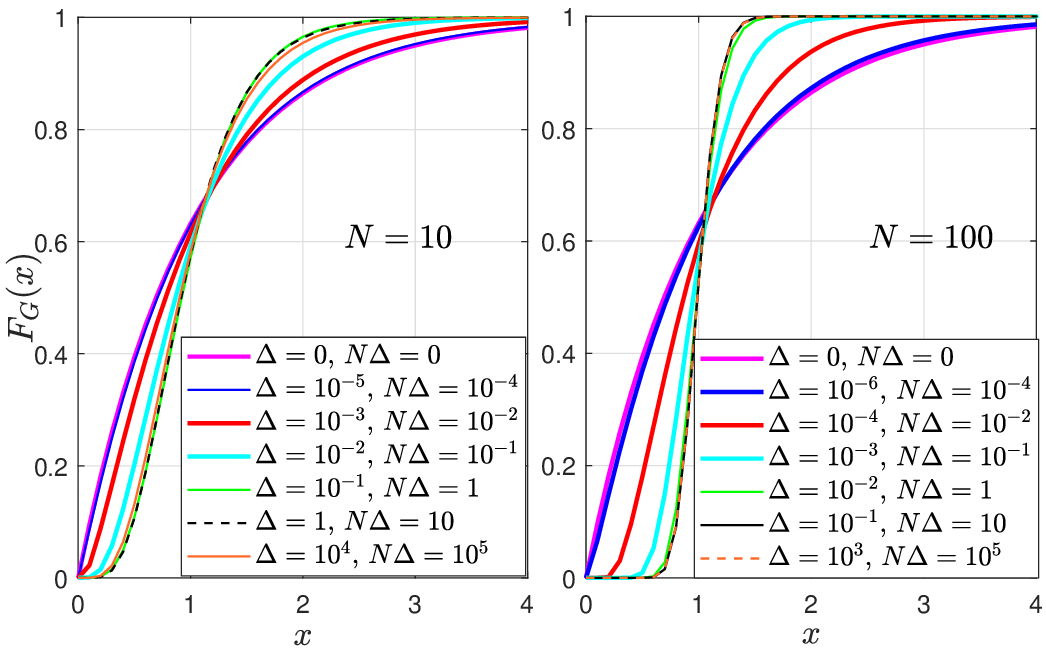}
\subcaption{ {$\mu=1$ (Rayleigh fading)}}\label{cdf_Y2_Rayl}
\end{minipage}
\begin{minipage}[htb]{\linewidth}
\centering\includegraphics[width=0.95\linewidth]{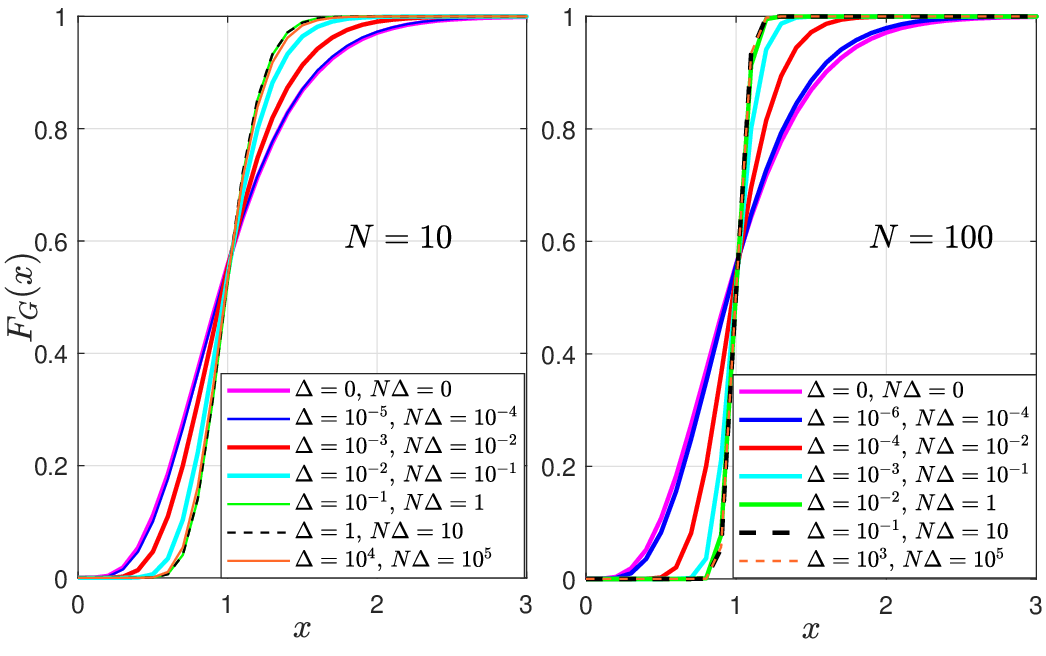}
\subcaption{$\mu=5$}\label{cdf_Y2_Nak}
\end{minipage}
\caption{The simulated CDF of $G$ for different $\Delta$ and $N$.}\label{Y2_tildebDiff}
\end{figure}

The ECPC $G$ includes both the impact of the direct link and the indirect link. Fig. \ref{Y2_tildebDiff} is a plot of the simulated CDF of $G$ for $N=10$ and $100$ using different $\Delta$. {Each simulation-based CDF curve of $G$ {in Fig. \ref{Y2_tildebDiff}} is obtained by generating $10^6$ realizations of the {Nakagami} RVs $g_0$, $g_{i,1}$, $g_{i,2}$ {with parameters $\mu$ and $\omega=1$} for $1 \leq i \leq N$ and using \eqref{Y} to obtain $10^6$ realizations of $\tilde{G}$. {Since} $G$ is the normalized version of $\tilde{G}$, the realizations of $G$ are obtained by scaling accordingly}. {It is observed that
\begin{itemize}
\item When $N\Delta \leq 10^{-4}$ {or $N\Delta \geq 1$}: the distribution of $G$ remains roughly unchanged. {We denote these ranges of smaller and larger $N\Delta$ values using $\chi=\rm s$ and $\chi=\rm l$ for small and large, respectively.}
\item When $10^{-4} < N\Delta < 1 $: the distribution of $G$ varies with $N\Delta$. {We denote this range of medium $N\Delta$ values using $\chi=\rm m$ for medium.}
\end{itemize}} 
Due to this dependence on $N\Delta$, {rather than just $\Delta$,} we observe in Fig. \ref{Y2_tildebDiff} that the impact of {varying $\Delta$, in the regime where the distribution of $G$ varies (i.e., $ 10^{-4} < N\Delta< 1$),} on the distribution of $G$ is greater for larger $N$ than in the case of smaller $N$. {This occurs because of the channel hardening effect, where the fading RV tends to a deterministic constant as $N$ grows \cite{IRS_SG2}; from Fig. \ref{Y2_tildebDiff} we observe that channel hardening occurs not just with growing $N$ but $N\Delta$.}  

Fig. \ref{varY2} is a plot of the simulated variance of $G$ vs. $N$ for different values of $\Delta$. {For each value of $N$ and $\Delta$, the simulation-based variance {in Fig. \ref{varY2}} is obtained by generating $10^6$ realizations of the {Nakagami} RVs $g_0$, $g_{i,1}$, $g_{i,2}$ {with parameters $\mu$ and $\omega=1$} for $1 \leq i \leq N$ and using \eqref{Y} to obtain $\tilde{G}$, followed by scaling to make them the unit-mean $G$ from which the variance is then calculated.} {We first observe from the figure that {while the variance is constant for all $N$ and equal to $1/\mu$ (corresponding to the variance of $g_0^2$) when $\Delta=0$,} the variance decreases with both $N$ and $\Delta$. This corroborates that both $N$ and $\Delta$ have an impact on channel hardening. We also observe that} the variance of $G$ decays at a rate lower than {${1}/{(\mu N)}$}. More precisely, {$1/{(\mu N^{0.75})}$ is a good approximate for larger $\Delta$ values, $1/\mu$ is a good approximate for smaller $\Delta$ values and $1/{(\mu N^{0.25})}$ is a good approximate for medium $\Delta$ values. }






In the following we introduce and motivate {some} promising candidate distributions to approximate the CDF of $G$. {We use $\chi \in \{\rm s,m,l\}$ to denote the range of $N\Delta$ the approximation will be used in and $\zeta \in \{\rm Rayl, Nak \}$ to denote whether the fading on the D\&I links is Rayleigh {($\mu=1$) or Nakagami with $\mu>1$}.} First, using the fact that $G$ is unit-mean {and unit-variance}, we propose the: 

\subsubsection*{Exponential Approximation for $G$}
The RV $G$ is approximated using a unit mean exponential RV denoted by {$G_{\rm exp}^{\chi,\zeta}$}, i.e.,
{ \begin{align}
{F_{G_{\rm exp}^{\chi,\zeta}}(y) } = 1-\exp\left(-  y \right), \label{cdfExp}
\end{align} }


\begin{figure}
\begin{minipage}[htb]{\linewidth}
\centering\includegraphics[width=0.8\linewidth]{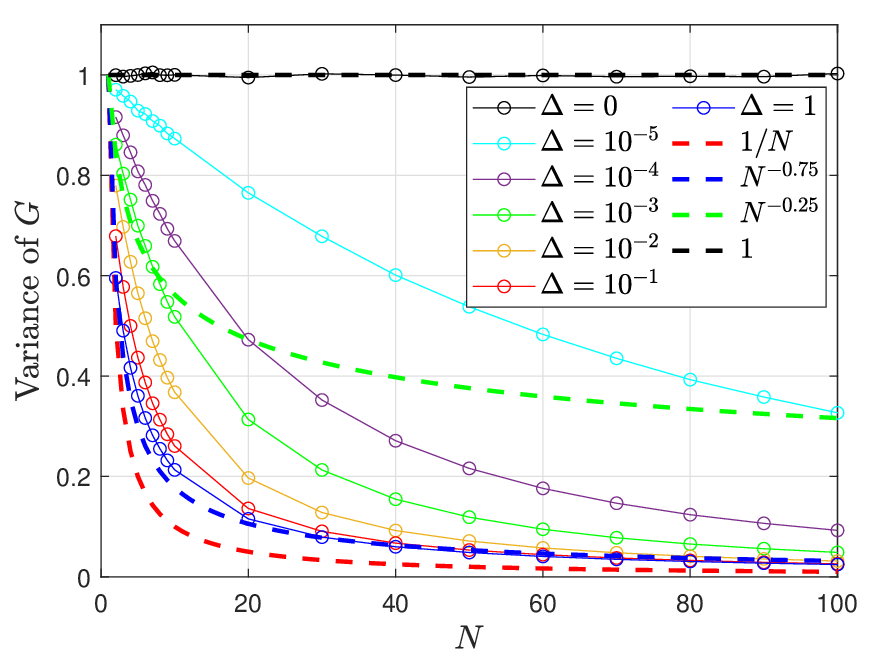}
\subcaption{{$\mu=1$ (Rayleigh fading)}}\label{varY2_Rayl}
\end{minipage}
\begin{minipage}[htb]{\linewidth}
\centering\includegraphics[width=0.8\linewidth]{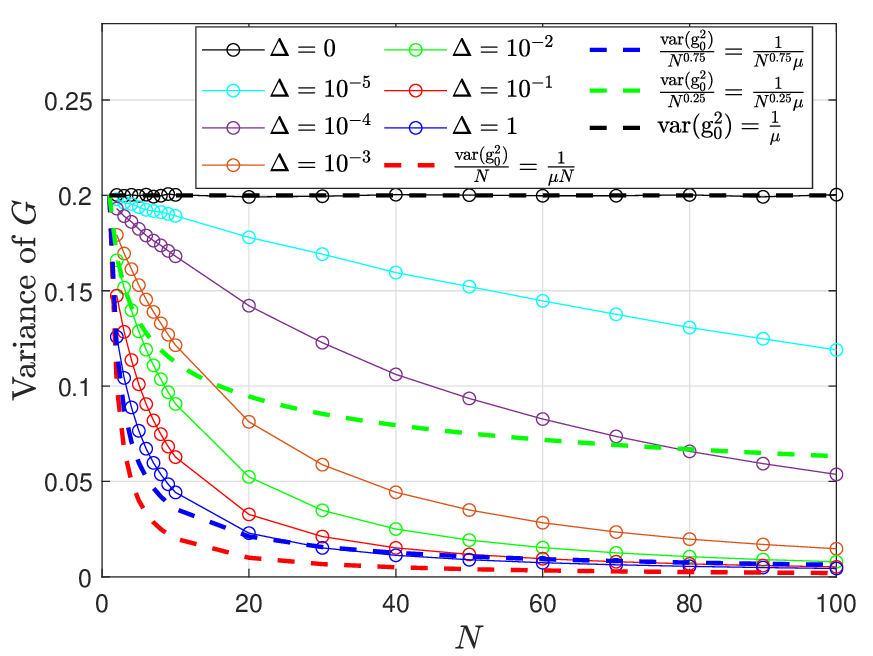}
\subcaption{{$\mu=5$}}\label{varY2_Nak}
\end{minipage}
\caption{Simulated variance of $G$ vs. $N$ for different $\Delta$. Also plotted are the ${\rm{var}(g_0^2)}{N^{-1}}$, ${\rm{var}(g_0^2)}{N^{-0.75}}$ and ${\rm{var}(g_0^2)}{N^{-0.25}}$ curves.}\label{varY2}
\end{figure}

{Second,} using the facts that $G$ is unit-mean and its variance is {approximated by $1/\bar{M}$, {for some $\bar{M}$}}, we propose the:

\subsubsection*{Gamma Approximation for $G$}
The RV $G$ is approximated using a gamma RV {$G_{\rm gam}^{\chi,\zeta}$} with parameters {$\bar{M}$ and $1/\bar{M}$} so that {$G_{\rm gam}^{\chi,\zeta}$} has unit mean and variance {$1/\bar{M}$, i.e., 
{   \begin{align}
F_{G_{\rm gam}^{\chi,\zeta}}(y)= \frac{1}{\Gamma(\bar{M})} \gamma \left(\bar{M}, \bar{M} y \right). \label{cdfGamma}
\end{align} } }
{We propose the following approximations for $G$ when
\begin{itemize}
\item $N\Delta \leq 10^{-4}$, i.e., $\chi=\rm s$: According to Fig. \ref{varY2}, the variance of $G$ in this scenario can be approximated as $1/\mu$. Accordingly, in the case of {$\zeta=\rm Nak$}, we approximate $G$ with $G_{\rm gam}^{\rm s, Nak}$ using $\bar{M}=\mu$. In the case of {$\zeta=\rm Rayl$}, $\mu=1$ and the gamma distribution becomes the exponential distribution; we thus approximate $G$ with $G_{\rm exp}^{\rm s, Rayl}$. 
\item $10^{-4} < N\Delta < 1$, i.e., $\chi=\rm m$: According to Fig. \ref{varY2}, the variance of $G$ in this scenario can be approximated as $N^{-0.25}/\mu$. Accordingly, in the case of {$\zeta=\rm Nak$ ($\rm \zeta=Rayl$)}, we approximate $G$ with $G_{\rm gam}^{\rm m, Nak}$ ($G_{\rm gam}^{\rm m, Rayl}$) using $\bar{M}=N^{0.25}\mu$ ($\bar{M}=N^{0.25}$).
\item $N\Delta \geq 1$, i.e., $\chi=\rm l$: According to Fig. \ref{varY2}, the variance of $G$ in this scenario can be approximated as $N^{-0.75}/\mu$. Accordingly, in the case of {$\zeta=\rm Nak$ ($\rm \zeta=Rayl$)}, we approximate $G$ with $G_{\rm gam}^{\rm l, Nak}$ ($G_{\rm gam}^{\rm l, Rayl}$) using $\bar{M}=N^{0.75}\mu$ ($\bar{M}=N^{0.75}$).
\end{itemize}
}

%

%



Fig. \ref{Y2_diffTildeB_diffApproxs} is a plot of some of the simulated CDFs of $G$ using some of the $\Delta$ in Fig. \ref{Y2_tildebDiff} and the analytical CDFs for the approximations proposed above for $N=10$ and $100$. {We observe in the figure that the CDF of $G$ closely matches the CDF of the approximations proposed above in each range of $N\Delta$ closely.} Unlike the CDF of $G_{\rm exp}^{\chi, \zeta}$, the CDF of $G_{\rm gam}^{\chi, \zeta}$, however, does not offer good tractability when it comes to obtaining the statistics of the SIR {for the case where the interferers form a PPP, which our model will be narrowed down to (cf. Sec \ref{sec3b})}. {The Erlang RV, on the other hand, offers better tractability when it comes to obtaining the statistics of the SIR and with carefully chosen parameters can approximate the gamma RV well. {Third,} we thus propose using an Erlang RV to approximate $G_{\rm gam}^{\chi, \zeta}$.}

\subsubsection*{Erlang Approximation for $G$}
The RV $G$ is approximated using an Erlang RV {$G_{\rm Erl}^{\chi, \zeta}$ with parameters $M=\text{round} \left( \bar{M} \right)$} and $1/M$ so that $G_{\rm Erl}^{\chi, \zeta}$ has unit mean and variance $1/M$, i.e., 
 \begin{align}
F_{G_{\rm Erl}^{\chi, \zeta}}(y)= 1- \sum_{k=0}^{M-1} \frac{1}{k!} \exp(-My) (My)^k . \label{cdfErl}
\end{align} 
In Fig. \ref{Y2_diffTildeB_diffApproxs} we observe that the CDF of $G_{\rm Erl}^{\chi, \zeta}$ is an excellent match for the CDF of $G_{\rm gam}^{\chi, \zeta}$. 

{\textbf{\emph{Remark 3:}} In the mid-range of $N\Delta$, i.e., $\chi= \rm m$, we use $\bar{M}=N^{0.25}\mu$. {This is done} so that the CDF of $G_{\rm gam}^{m,\zeta}$ (and therefore $G_{\rm Erl}^{m,\zeta}$) closely matches the CDF of $G$ when $N\Delta=10^{-2}$, which is in the middle of the mid-range and approximates the range well. However, when $\chi= \rm m$, using $\bar{M}=N^{\frac{3}{4} + \frac{\log_{10}(N\Delta)}{4}}\mu$ results in a good match with the CDF of $G$ for any $N\Delta$ value.}

\begin{table*}[htb]
  \centering
  \renewcommand{\arraystretch}{1.2}
  \begin{tabular}{|p{2.5cm} |p{1cm}|c|c|c|c|}
    \hline
    \multirow{2}{2.5cm}{$N\Delta$} & \multirow{2}{1cm}{$\chi$} & \multicolumn{2}{c|}{\textbf{Rayleigh fading ($\mu=1$): {$\zeta={\rm Rayl}$}} } &  \multicolumn{2}{c|}{\textbf{\textbf{Nakagami fading {with $\mu>1$: $\zeta={\rm Nak}$}}}} \\
    \cline{3-6}
    &   & \textbf{Approximation of $G$} & \textbf{$M$} &  \textbf{Approximation of $G$} & \textbf {$M$}\\
    \hline
    $N\Delta \leq 10^{-4}$ & {\rm s} & $G_{\rm exp}^{\rm s, Rayl}$ & N/A & $G_{\rm Erl}^{\rm s, Nak}$ & $\text{round}(\mu)$ \\ \hline
    $10^{-4} < N\Delta < 1 $ & {\rm m} & $G_{\rm Erl}^{\rm m, Rayl}$ & $\text{round}(N^{0.25})$ & $G_{\rm Erl}^{\rm m, Nak}$ & $\text{round}(N^{0.25}\mu)$ \\ \hline
    $ N\Delta \geq 1 $ & {\rm l} & $G_{\rm Erl}^{\rm l, Rayl}$ & $\text{round}(N^{0.75})$ & $G_{\rm Erl}^{\rm l, Nak}$ & $\text{round}(N^{0.75}\mu)$  \\ \hline
  \end{tabular}
  \caption{Range of $N\Delta$ and the corresponding approximation that closely matches the CDF of $G$.}\label{NDeltaTable}
\end{table*}


While we have only plotted the CDFs in Fig. \ref{Y2_diffTildeB_diffApproxs} for $N=10$ and $100$ for brevity, these results were found to hold for general $N$. {Since $G_{\rm gam}^{\chi, \zeta}$ does not lead to tractable expressions for the SIR CDF, in the remainder of this work, we will not discuss it further and focus on} the CDFs of 1) $G_{\rm exp}^{\rm s, Rayl}$ and {2) $G_{\rm Erl}^{\chi, \zeta}$}. Table \ref{NDeltaTable} summarizes the ranges of $N\Delta$ and the corresponding approximation for the CDF of $G$ {with its parameters where appropriate, for the cases of {Nakagami fading with $\mu=1$ (Rayleigh) and $\mu>1$} on the D\&I links.}

{While prior work has focused solely on the large-$N$ case where the gamma distribution {with a fixed parameter} is often the natural approximation, we {explore} the parameter space comprehensively. Further, we have shown that $N\Delta$ is relevant for determining the approximate statistics of $G$ --- even if $N$ is large, $N\Delta$ could still be small. Additionally, {for both {$\zeta={\rm Rayl}$ and $\zeta={\rm Nak}$,}} while it is clear that as $N \to \infty$, the gamma distribution {$G_{\rm gam}^{\rm l, \zeta}$, and therefore $G_{\rm Erl}^{\rm l, \zeta}$,} is a good approximation for the distribution of $G$, we observe that for each $N$, $G_{\rm Erl}^{\rm l, \zeta}$ is also the limiting distribution as $\Delta \to \infty$. }


{Also note that while the approximations for the distribution of $G$ proposed above are for the IRS-assisted case ($N>0$), in the no-IRS scenario ($N=0$), the distribution of $G$ is known exactly as {$G=\|g_0\|^2=h_0$, and {$h_0 \sim \text{gamma}(\mu,1/\mu)$ (for $\mu=1$, this becomes $h_0 \sim \exp(1)$)}.} The proposed {approximation for the} CDF of $G$ for $N>0$ {when $\chi= \rm s$} in \eqref{cdfGamma} {with $\zeta= \rm Nak$ (in \eqref{cdfExp} with $\zeta= \rm Rayl$)} is designed to have the same statistics as the exact CDF of $G$ in the no-IRS case.}

\begin{figure}
\begin{minipage}[htb]{\linewidth}
\centering\includegraphics[width=0.95\linewidth]{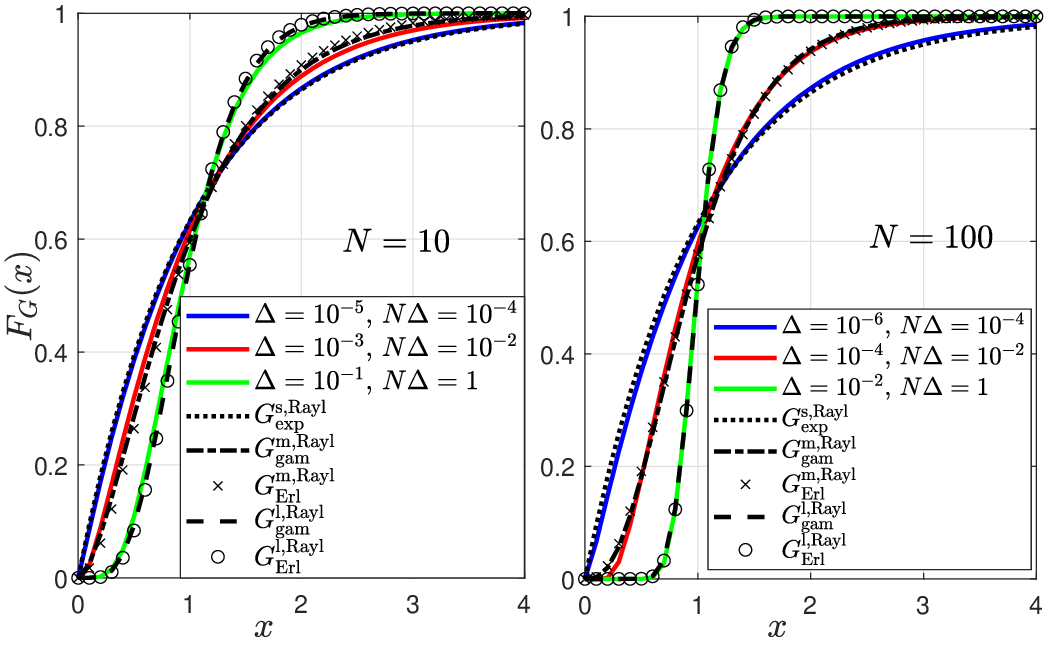}
\subcaption{{$\mu=1$ (Rayleigh fading)}}
\end{minipage}\;\;
\begin{minipage}[htb]{\linewidth}
\centering\includegraphics[width=0.95\linewidth]{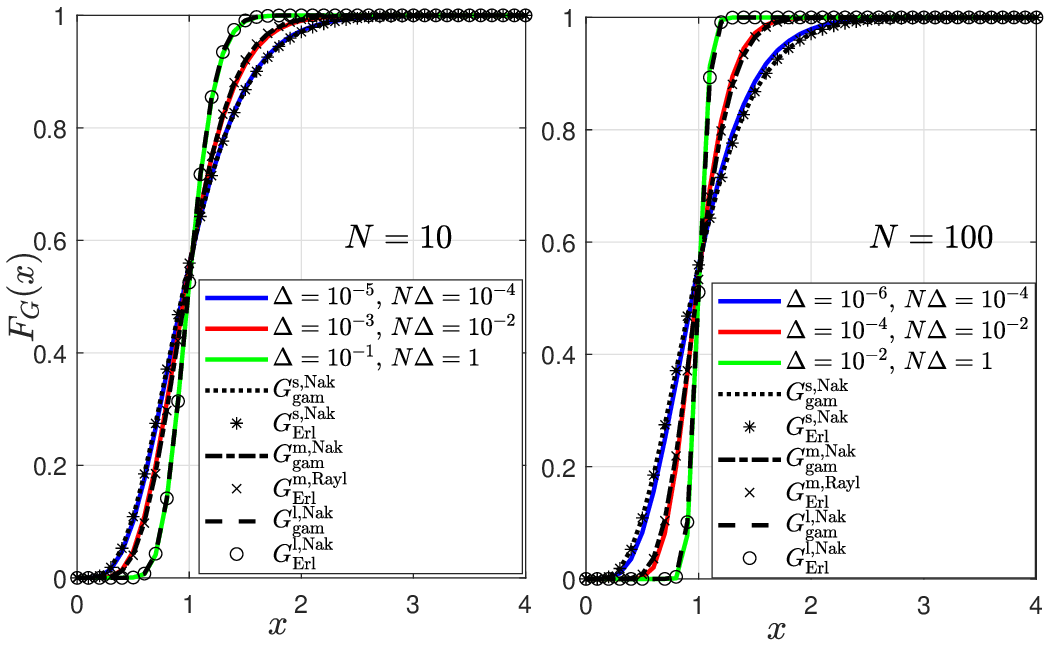}
\subcaption{{$\mu=5$}}
\end{minipage}
\caption{The CDF of $G$ for different $\Delta$ and the approximations for the CDF of $G$ {in \eqref{cdfExp}, \eqref{cdfGamma} and \eqref{cdfErl}}.}\label{Y2_diffTildeB_diffApproxs}
\end{figure}

%
%
%
%
%
%
%


\subsection{Point Process of BSs}\label{sec3b}
In the remainder of this work we assume the BS point process $\Phi$ is a homogeneous PPP with intensity $\lambda$. To the network we add a BS at the origin $\textbf{o}$, which, under expectation over $\Phi$, becomes the tBS. Since $\Phi$ does not include the BS at $\textbf{o}$, the set of interfering BSs for the tUE is $\Phi$. The tUE lies at a distance $R_0$ from the tBS at $\textbf{o}$. As UEs are distributed uniformly at random in the cells and the BSs are distributed according to a PPP, 
\begin{align}
f_{R_0}(x) = 2\pi q \lambda x \exp(-\pi q \lambda x^2), \;\;\;\; x \geq 0, \label{f_R0}
\end{align} 
where $q=9/7$ is the correction factor {relative to the standard empty space function of the PPP} due to the fact that the tUE lies in the typical cell, not in the 0-cell \cite{typicalCell,mh_ue_PP}.

\subsection{On the Laplace Transform (LT) of the Interference}
Before delving into the analysis of the CCDF of the SIR, we {study} the LT of the interference. {Recall that we assume Rayleigh fading on all the interfering links.}

When the UEs form a stationary point process independent of $\Phi$, the interferers at the tUE form a homogeneous PPP with density $\lambda$ outside of $b(\textbf{u},R_0)$ \cite{myD2DFD}. The corresponding LT of interference conditioned on $R_0$ is
{ \begin{align}
&\mathcal{L}_{I \mid R_0}(s)\!\approx\!\exp \left(\!   \frac{-2 \pi \lambda s \ell_0^{\eta}}{(\eta\!-\!2) R_0^{\eta-2}} { }_2F_1  \! \left(\! 1,1 \! - \!  \delta; 2 \!- \! \delta; \! \frac{-s \ell_0^{\eta}}{R_0^{\eta}} \! \right) \!  \right) \label{L_Iold}\\
&\stackrel{\eta=4}= \exp \left({-\pi \lambda \sqrt{s} \ell_0^2 \tan^{-1} \left({\sqrt{s}\ell_0^2 R_0^{-2} } \right)} \right)  .
\end{align}
}
It was shown in \cite{typicalCell}, however, that when the UEs are distributed uniformly at random and independently in each cell (as done in this work), {the BS and user point processes are dependent}. {It was shown that} from the perspective of the tUE, the point process of interferers exhibits a clustering effect at distances slightly larger than the link distance $R_0$. Approximating this point process using a homogeneous PPP with density $\lambda$ outside of $b(\textbf{u},R_0)$ thus leads to an underestimation of the interference. {Let us denote the distance between the tUE at $\textbf{u}$ and the dominant interferer, which is the interferer nearest to the tUE, by $R_{\rm d}=\arg \min_{\textbf{x} \in \Phi} \|\textbf{x} - \textbf{u} \|$.} While the exact characterization of the point process of interferers is complex, it was shown in \cite{typicalCell} that {explicitly considering the interference from} the dominant interferer {and the approximate point process of interferers with density $\lambda$ outside of $b(\textbf{u},R_{\rm d})$} is a much better approximation than just a homogeneous PPP with density $\lambda$ outside of $b(\textbf{u},R_0)$, {resulting} in a very good estimate of the interference. The point process of interferers is therefore modeled as the union of the interferer at distance $R_{\rm d}$ from the tUE and a homogeneous PPP with intensity $\lambda$ outside of $b(\textbf{u}, R_{\rm d})$. The resultant interference $I$ at the tUE is 
{  \begin{align}
I =\sum_{\substack{\textbf{x}\in\Phi\\ \|\textbf{y}\|>R_{\rm d}}}   {\|\textbf{y} \|}^{-\eta} \ell_0^{\eta} h_{\textbf{y}} + R_{\rm d}^{-\eta} \ell_0^{\eta} {h_{\rm d}} , \label{I_Rd}
\end{align} }
{where $h_{\rm d}$ is the channel power coefficient from the interfering BS located a distance $R_{\rm d}$ from the tUE and the tUE.}

\textbf{\emph{Lemma 2:}} The LT of the interference at the tUE conditioned on $R_{\rm d}$ is approximated as
{\small \begin{align}
&\mathcal{L}_{I \mid R_{\rm d}}(s)\!\approx\!\exp \left(\!   \frac{-2 \pi \lambda s \ell_0^{\eta}}{(\eta\!-\!2) R_{\rm d}^{\eta-2}} { }_2F_1  \! \left(\! 1,1 \! - \!  \delta; 2 \!- \! \delta; \! \frac{-s \ell_0^{\eta}}{R_{\rm d}^{\eta}} \! \right) \!  \right) \! \frac{1}{1 \!+ \! s \ell_0^{\eta} R_{\rm d}^{-\eta} } \label{L_I}\\
&\stackrel{\eta=4}= \exp \left({-\pi \lambda \sqrt{s} \ell_0^2 \tan^{-1} \left({\sqrt{s}\ell_0^2 R_{\rm d}^{-2} } \right)} \right) \frac{1}{1 \!+ \! s \ell_0^{\eta} R_{\rm d}^{-\eta} } .
\end{align} }
\textbf{\emph{Proof:}} By applying \eqref{I_Rd} to the definition of the LT, we obtain two terms. Using the probability generating functional (PGFL) of the PPP and the fact that the channel power coefficients between the tUE and the interfering BSs are unit-mean exponential distributed, the first term in \eqref{L_I} is obtained. The second term of the LT, which comes from using the second term of \eqref{I_Rd}, is {$\mathbb{E}_{h_{\rm d}}\left[\exp \left( -s R_{\rm d}^{-\eta} \ell_0^{\eta} h \right)\right]$}. As {$h_{\rm d} \sim \exp(1)$,} the second term in \eqref{L_I} is obtained. \qed

From \cite{distDistrib,typicalCell}, the distribution of $R_{\rm d}$ conditioned on $R_0$, {due to the clustering effect at distances slightly larger than $R_0$,} is calculated as
{ \begin{align}
f_{R_{\rm d} \mid R_0}(r \mid R_0) = 2\pi\lambda q r \exp \left(-\pi \lambda q \left(r^2-R_0^2 \right) \right), \; r \geq R_0.
\end{align}   
}


\subsection{On the SIR Distribution}
The CCDF of the SIR expression in terms of $G$ was given in \eqref{ccdfSIR_Y2}. In this subsection we focus on the approximations of {the CDF of} $G$ that result in a tractable expression for the CCDF of the SIR, {namely, $G_{\rm exp}^{\rm s, Rayl}$ and $G_{\rm Erl}^{\chi, \zeta}$.}

{For $N>0$,} if the CDF of $G_{\rm exp}^{\rm s, Rayl}$ {in \eqref{cdfExp}} is used to {approximate} $G$, the CCDF of the SIR is
{\small \begin{align}
\mathbb{P}({\rm SIR}>\theta)=\mathbb{E}\left[ \bar{F}_{G_{\rm exp}^{\rm s, Rayl}} \left( \frac{\theta R_0^{\eta} \ell_0^{-\eta} }{ \mathbb{E}[\tilde{G}\mid \Delta]} I \right) \right] =\mathbb{E}\left[  \mathcal{L}_{I \mid R_{\rm d}} \left( \frac{ \theta R_0^{\eta} \ell_0^{-\eta}}{ \mathbb{E}[\tilde{G}\mid \Delta]} \right) \right]. \label{ccdfSIR_Y2Exp}
\end{align} }

{{For $N>0$,} if the CDF of $G_{\rm Erl}^{\chi, \zeta}$ {in \eqref{cdfErl}} is used to approximate $G$, the CCDF of the SIR is
{\small \begin{align}
&\mathbb{P}({\rm SIR}>\theta)=\mathbb{E}\left[ \bar{F}_{G_{\rm Erl}^{\chi, \zeta}} \left( \frac{\theta R_0^{\eta} \ell_0^{-\eta}}{ \mathbb{E}[\tilde{G}\mid \Delta]} I \right) \right] \nonumber\\
&= \sum_{k=0}^{M-1} \frac{1}{k!} \mathbb{E}\left[  \exp \left(- \frac{M \theta R_0^{\eta} \ell_0^{-\eta}}{ \mathbb{E}[\tilde{G}\mid \Delta]} I \right) \left( \frac{M \theta R_0^{\eta} \ell_0^{-\eta}}{ \mathbb{E}[\tilde{G}\mid \Delta]} I \right)^k \right] \nonumber \\
&= \sum_{k=0}^{M-1} \frac{1}{k!} \mathbb{E}\left[  \mathcal{L}_{I \mid R_{\rm d}}^{(k)} \left( \frac{M \theta R_0^{\eta} \ell_0^{-\eta}}{ \mathbb{E}[\tilde{G}\mid \Delta]} \right) \left( \frac{M \theta R_0^{\eta} \ell_0^{-\eta}}{ \mathbb{E}[\tilde{G}\mid \Delta]}  \right)^k \right], \label{ccdfSIR_Y2Erl}
\end{align} }

where $\mathcal{L}_{I \mid R_{\rm d}}^{(k)}(s)$ denotes the $k^{\rm th}$ derivative of the LT $\mathcal{L}_{I \mid R_{\rm d}}(s)$ w.r.t. 
$s$. 
Computing the CCDF of the SIR via the approach in \cite[Theorem 2]{mh2} using the LT of interference conditioned on $R_{\rm d}$ is computationally challenging. We thus resort to using the LT of interference conditioned on $R_0$ which underestimates the interference but results in tractable expressions for the SIR CCDF and trade some accuracy for tractability again. {We account for the clustering of interferers outside of $R_0$ (as shown in \cite[Fig.~1 (left)]{typicalCell}) by increasing the interferer density to $q\lambda$.} Along the lines of \cite[Corollary 1]{mh2}, the SIR CCDF is 

{\small \begin{align}
\mathbb{P}({\rm SIR}>\theta) &{= \sum_{k=0}^{M-1} \frac{1}{k!} \mathbb{E}\left[  \mathcal{L}_{I \mid R_{\rm 0}}^{(k)} \left( \frac{M \theta R_0^{\eta} \ell_0^{-\eta}}{ \mathbb{E}[\tilde{G}\mid \Delta]} \right) \left( \frac{M \theta R_0^{\eta} \ell_0^{-\eta}}{ \mathbb{E}[\tilde{G}\mid \Delta]}  \right)^k \right]} \label{ccdfSIR_Y2ErlUSE0} \\
& = \| {\textbf{C}_M}^{-1}\|_1 \label{ccdfSIR_Y2ErlUSE}
\end{align} }
where $\textbf{C}_M$ is the $M \times M$ lower triangular Toeplitz matrix 
{\footnotesize
\begin{align}
\begin{bmatrix}
c_0 &  & \\
c_1 & c_0 & \\
\vdots & \vdots & \ddots \\
c_{M-1} & c_{M-2} &\hdots & c_0  
\end{bmatrix}
\end{align} }
with 
{\small \begin{align} 
c_k \approx \left(\! \frac{\theta M}{\mu_{\tilde{G}}} \!\right)^k \frac{(1)_k (-\delta)_k  {}_2F_1 \left(k\!+\!1, k\!-\!\delta; k\!-\!\delta\!+\!1; \frac{-\theta M}{\mu_{\tilde{G}}} \right)}{k!  \; (1\!-\!\delta)_k} , \label{c_k}
\end{align} }
and $\| .  \|_1$ denotes the $l_1$ induced matrix norm. In \eqref{c_k}, $\mu_{\tilde{G}}=\mathbb{E}_{\Delta}[\mathbb{E}[\tilde{G}\mid \Delta]]$. Note that in \cite{mh2}, this approach led to an exact result. {In our scenario, however, the derivative of the LT in \eqref{ccdfSIR_Y2Erl} and \eqref{ccdfSIR_Y2ErlUSE0} is evaluated at $\frac{M \theta R_0^{\eta} \ell_0^{-\eta}}{ \mathbb{E}[\tilde{G}\mid \Delta]}$ which includes the signal amplification term $\mathbb{E}[\tilde{G}\mid \Delta]$. Since $\mathbb{E}[\tilde{G}\mid \Delta]$ is a function of $\Delta$, and therefore of $R_0$ and $R_1$, in \eqref{ccdfSIR_Y2ErlUSE} we resort to approximating this by using $\mu_{\tilde{G}}$, the mean of $\mathbb{E}[\tilde{G}\mid \Delta]$ w.r.t. $R_0$ and $R_1$.} This approximation is justified as we will see in the next section that the $\Delta$ for Model I is concentrated around its mean and the $\Delta$ in Model II is a constant independent of $R_0$ and $R_1$ (making the approximation in \eqref{ccdfSIR_Y2ErlUSE} an exact expression). For the remainder of this work, we compute the CCDF of the SIR based on the CDF of $G_{\rm Erl}^{\chi, \zeta}$ using \eqref{ccdfSIR_Y2ErlUSE}.

}


%


For {$N=0$}, the case without IRS assistance, $\mathbb{E}[\tilde{G}\mid \Delta]=1$ in \eqref{sinr_gen0}. {With Rayleigh fading, $\tilde{G}=G=h_0 \sim \exp(1)$ and the SIR CCDF is
{\begin{align}
&\mathbb{P}({\rm SIR}>\theta)=\mathbb{E}\left[ \bar{F}_{h_0} \left( {\theta R_0^{\eta} \ell_0^{-\eta}} I \right) \right] =\mathbb{E} \left[   \mathcal{L}_{I \mid R_{\rm d}} \left( { \theta R_0^{\eta} \ell_0^{-\eta}} \right) \right]. \label{ccdfSIR_noIRS}
\end{align} }
For $N=0$ with Nakagami fading, $\tilde{G}=G=h_0 \sim \text{gamma}(\mu,1/\mu)$ and the SIR CCDF {in the case of $\mu>1$} is calculated using \eqref{ccdfSIR_Y2ErlUSE} with $\mu_{\bar{G}}=1$ and $M=\text{round}(\mu)$ in \eqref{c_k}.}

\textbf{\emph{Remark 4:}} {For Rayleigh fading,} while the CDF of $G_{\rm exp}^{\rm s, Rayl}$ is equal to that of $G$ without IRS assistance, the fact that the approximations of $G$, including $G_{\rm exp}^{\rm s, Rayl}$, are unit-mean results in the $\mathbb{E}[\tilde{G}\mid \Delta]$ term in \eqref{ccdfSIR_Y2} and therefore in \eqref{ccdfSIR_Y2Exp}. The CCDF of the SIR in the IRS-assisted cases using the $G_{\rm exp}^{\rm s, Rayl}$ approximation is thus not equivalent to the CCDF of the SIR with no IRS (cf. \eqref{ccdfSIR_noIRS}). {Similarly, {for Nakagami fading with $\mu>1$,} the SIR CCDF with no IRS is not equivalent to the SIR CCDF using $G_{\rm Erl}^{\rm s, Nak}$ as $\mu_{\bar{G}}$ may not be $1$ in the latter.}

{For a given SIR threshold $\theta$, corresponding to a transmission rate $\log(1+\theta)$, the throughput of the tUE is: $\mathcal{T}= \mathbb{P}({\rm SIR}>\theta) \log(1+\theta).$

\subsection{Diversity Analysis}

The diversity gain, a metric used to measure the reliability of wireless communication schemes under fading, was defined in \cite[Def. 3]{mh_diversityJrnl} for the interference-limited case as 
{ \begin{align}
d = \lim_{\substack{\theta \to 0}} \frac{\log \mathbb{P}({\rm SIR} < \theta)}{ \log \theta}. \label{diversityDef}
\end{align}  }
{This means that $\mathbb{P}({\rm SIR} < \theta) = \Theta(\theta^d)$, $\theta \to 0$.}



The SIR CDF, when using the CDF of $G_{\rm exp}^{\rm s, Rayl}$ and in the no-IRS scenario {with Rayleigh fading}, can be written in the following general form:
{\small \begin{align}
& \mathbb{P}({\rm SIR} <  \theta) =  \mathbb{E}  \left[  \left( 1 - {\exp \left(  {- \theta  \mu_1 \gamma_1}  I \right)}  \right)  \right] , \label{cdfSIR_gen}
\end{align} }
where in addition to the distances $R_0$, $R_1$ and $R_{\rm d}$, the expectation  in \eqref{cdfSIR_gen} is also over the interference $I$. Further, $\mu_1=R_0^{\eta} \ell_0^{-\eta}$ {and $\gamma_1=\frac{1}{\mathbb{E}[\tilde{G} \mid \Delta]}$ when using $G_{\rm exp}^{\rm s, Rayl}$, while $\gamma_1=1$ in the the no-IRS scenario with Rayleigh fading.}

{\textbf{\emph{Lemma 3:}} For the SIR CDF in \eqref{cdfSIR_gen}, the diversity gain{\footnote{{The diversity gain in interference-limited settings is a non-trivial matter due to the correlation in the interference, as analyzed in \cite{mhDivLoss}.}}} is lower bounded by 1, i.e., $d\geq 1$.}


\textbf{\emph{Proof:}} See Appendix \ref{L3proof}.

{The SIR CDF, when using the CDF of {$G_{\rm Erl}^{\rm \chi, \zeta}$}, is written as
{\small \begin{align}
& \mathbb{P}({\rm SIR} <  \theta) =  \mathbb{E}  \left[  1 - \sum_{k=0}^{M-1} \frac{1}{k!}\exp \left(  {- \theta  \gamma_E}  I \right) \left( \theta   \gamma_E I \right)^k  \right] , \label{cdfSIR_genErl}
\end{align} }
where $\gamma_E= \frac{M R_0^{\eta} \ell_0^{-\eta}}{\mathbb{E}[\tilde{G} \mid \Delta]}$ and in addition to the distances, the expectation  in \eqref{cdfSIR_genErl} is also over the interference $I$. }

{\textbf{\emph{Lemma 4:}} For the SIR CDF in \eqref{cdfSIR_genErl}, the diversity gain is {approximated} by $M$.}

{\textbf{\emph{Proof:}} See Appendix \ref{L4proof}. }

\textbf{\emph{Remark 5:}} {Due to the approximations involved, it is unclear whether the {result} in Lemma 4 is an upper or lower {bound}. From simulation results (cf. Figs. \ref{diversityVsN}, \ref{diversityVsN_Nak}), we find that for {$\zeta={\rm Rayl}$} it is a lower bound when $\chi= \rm s$ or $\rm m$ and is an upper bound when $\chi= \rm l$. For {$\zeta={\rm Nak}$}, it is an upper bound for all $\chi$. As increasing $N\Delta$ improves the diversity {for any $\mu$}: when $\chi = \rm l$, $d<\text{round}(\mu N^{0.75})$ holds for all $N\Delta$. For {$\zeta={\rm Nak}$}, the tighter $d<\text{round}(\mu N^{0.25})$ holds for $N\Delta <1$ and the tightest $d<\text{round}(\mu)$ holds for $N\Delta <10^{-4}$. For {$\zeta={\rm Rayl}$}, we have $d>1$ for $N\Delta <10^{-4}$ and the tighter $d > \text{round}(N^{0.25})$ when $N\Delta<1$.  } 


\section{Models for IRS Placement}
{Deploying IRSs completely at random can be inefficient as it is more likely to lead to scenarios with large $R_1$ and large $R_2$, in turn deterring $\Delta$.} In this section we propose two models for IRS placement {one of which is more practical and has a RV $\Delta$ and the other resulting in a deterministic $\Delta$}.

\begin{figure}[htb]
\begin{minipage}[htb]{\linewidth}
\centering\includegraphics[width=0.6\linewidth]{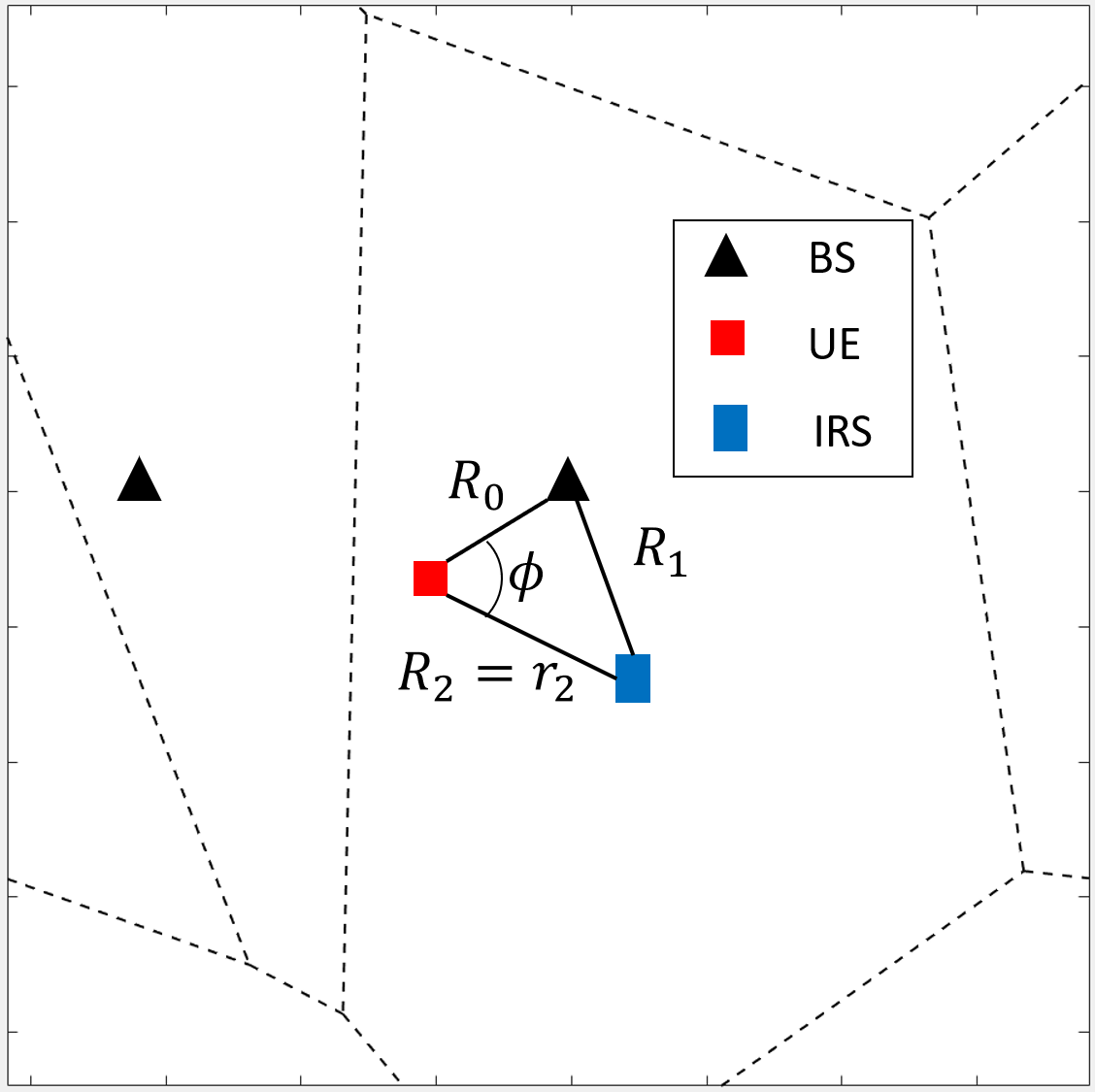}
\subcaption{{Model I: The distance $R_2=r_2$ is deterministic and the angle $\phi$ is shown.}}\label{model1}
\end{minipage}
\begin{minipage}[htb]{\linewidth}
\centering\includegraphics[width=0.6\linewidth]{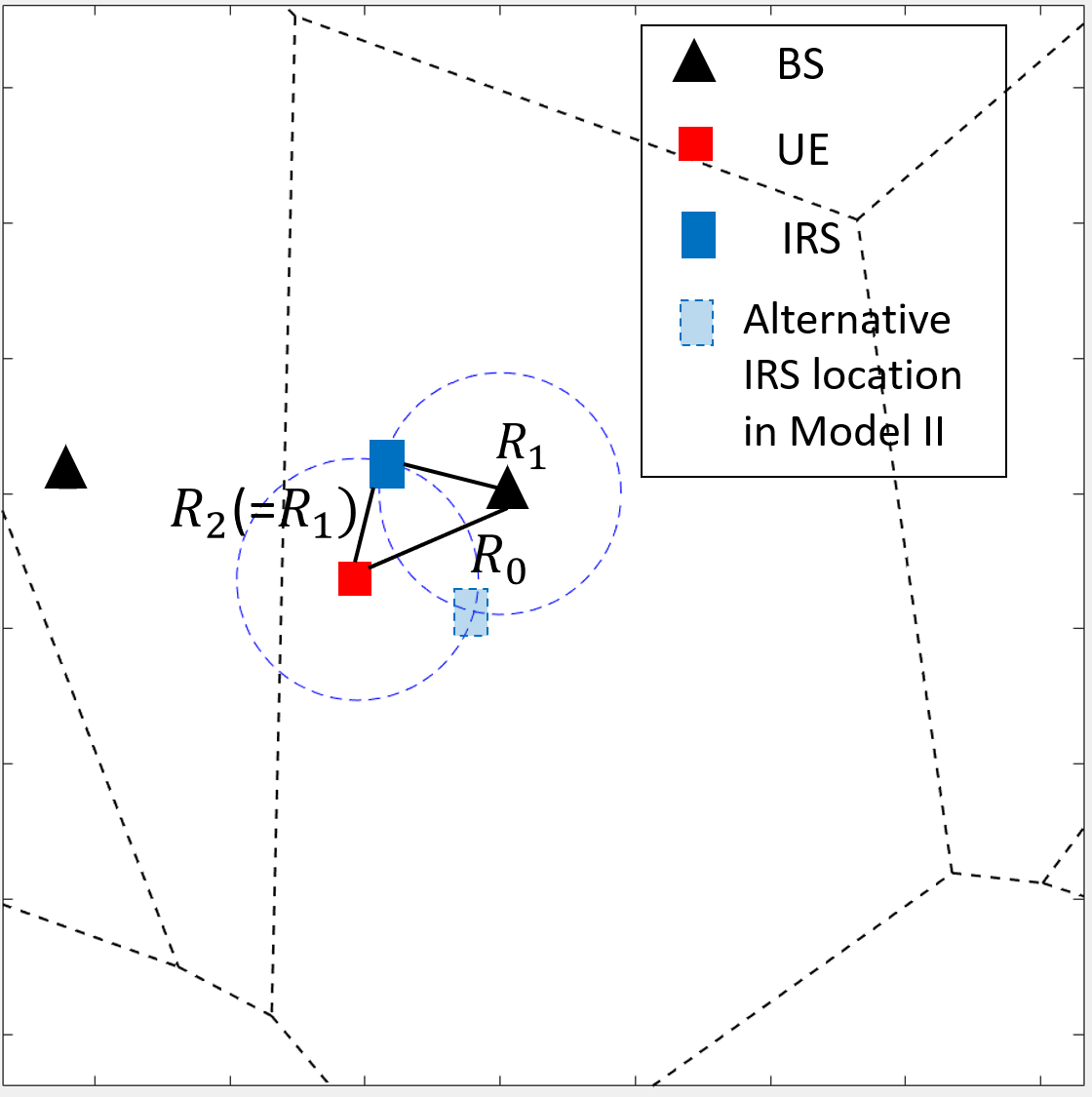}
\subcaption{Model II: The dashed circles centered at the tBS and the tUE each have radius $R_1=R_2$. As the tIRS is equidistant from the tBS and tUE, it can be located at one of the intersections of the two circles for a given $R_1$, as shown.}\label{model2}
\end{minipage}
\caption{Network realizations for Models I and II. {The tUE, tIRS and tBS in the typical cell are shown. Also seen is a neighboring interfering BS outside the typical cell.}}\label{models1and2}
\end{figure}

\subsection{Model I: {Random Direction IRS Deployment}}
{In this setup, we assume that $R_2$, the distance between the tIRS and tUE, {is deterministic and fixed to $r_2$} and that the tIRS is located in a random direction. The angle between the tUE-tIRS link and the tUE-tBS link, as shown in Fig. \ref{model1}, is denoted by $\phi$ and follows the distribution $f_{\phi}(u) =1/(2\pi)$, $0\leq u \leq 2\pi$. As a result, the distance between the tIRS and tBS at $\textbf{o}$, denoted by $R_1$, is
{ \begin{align}
R_1=\sqrt{R_0^2 + r_2^2 - 2 r_2 R_0 \cos \phi}. \label{R1}
\end{align}}

{\textbf{\emph{Proposition 1:}} The distribution of $R_1$ conditioned on $R_0$ in Model I, where $|R_0-r_2| \leq  R_1  \leq  R_0 + r_2$, is 
\begin{align}
&F_{R_1 \mid R_0}(x \mid R_0) = \frac{1}{2} - \frac{1}{\pi}  \sin^{-1} \left( \frac{R_0^2+r_2^2-x^2}{2 R_0 r_2} \right). \label{F_R1}  
\end{align}
\textbf{\emph{Proof:}} See Appendix \ref{P1proof}.

{\textbf{\emph{Remark 6:}} $F_{R_1 \mid R_0}(r \mid R_0)$ is unchanged if the values of $R_0$ and $r_2$ are interchanged as this means the positions of the BS and IRS in Fig. \ref{model1} have been swapped. 
\begin{figure}
\begin{minipage}[htb]{\linewidth}
\centering\includegraphics[width=0.95\linewidth]{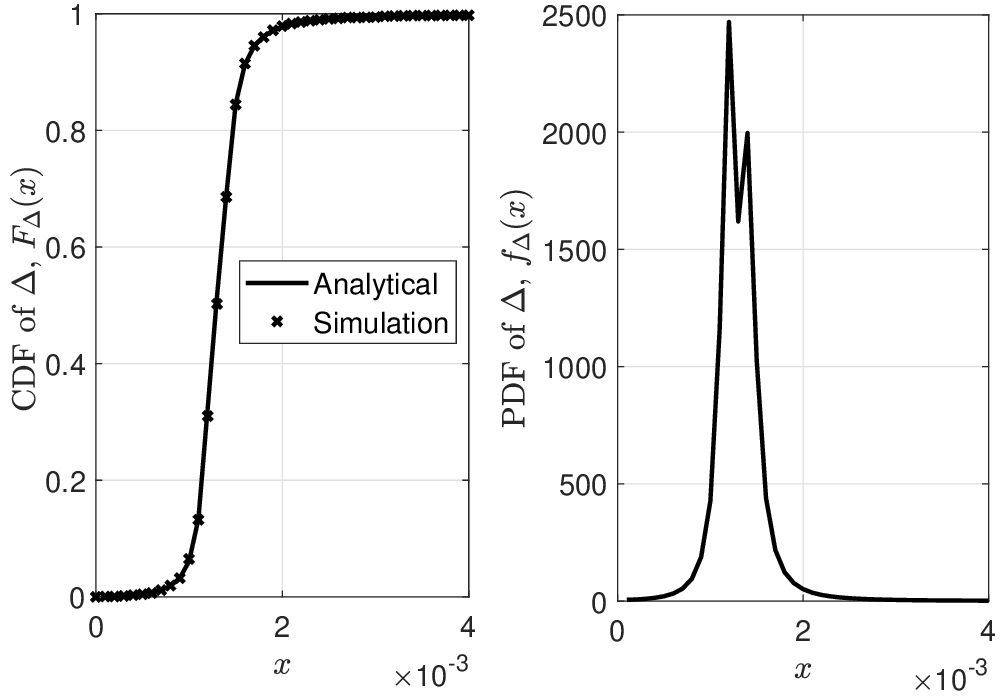}
\caption{The CDF and PDF of $\Delta$ in Model I from Corollary 1 {with $\lambda=10^{-5}$, {$r_2=1/(60\sqrt{\lambda})$} and $\ell_0=1$}.}\label{deltaStats}
\end{minipage}
\end{figure}

{The triangle parameter for Model I is $\Delta=\left(\frac{R_0 \ell_0}{R_1 r_2}\right)^{\eta}$.}
We can write the SIR for Model I as the SIR in \eqref{sinr_gen0} evaluated at $\Delta=\left(\frac{R_0 \ell_0}{R_1 r_2}\right)^{\eta}$.



Based on these, we derive the statistics of $\Delta$ for Model I.

\textbf{\emph{Corollary 1:}} The distribution of the triangle parameter $\Delta$ in Model I is
{\small {\begin{align}
&F_{\Delta}(x)=\frac{1}{2}  + \int_0^{\infty}  \frac{2  \lambda q y }{ e^{\!\pi \lambda q y^2}} \sin^{-1}  \Bigg( \frac{y^2 +r_2^2 - {y^2\ell_0^2}{ r_2^{-2} x^{\frac{-2}{\eta}} } }{2y r_2}  \Bigg)   dy, \; x \geq 0. \label{F_DeltaM1} 
\end{align}}}
\textbf{\emph{Proof:}} By definition of the CDF of $\Delta$ we have
{\begin{align*}
&F_{\Delta}(x)=\mathbb{P} \left(\! \left(\! \frac{R_0 \ell_0}{R_1 r_2} \!\right)^{\eta} < x \!\right) = \mathbb{E}_{R_0} \left[ 1 - F_{R_1 \mid R_0} \left( \frac{R_0 \ell_0}{r_2 x^{\frac{1}{\eta}}} \right) \right] .
\end{align*} }
{Using the CDF of $R_1$ in \eqref{F_R1} and the distribution of $R_0$ in \eqref{f_R0}, we obtain the CDF in \eqref{F_DeltaM1}. {The PDF can be obtained by taking the derivative of \eqref{F_DeltaM1}. \qed} 
}


{Fig. \ref{deltaStats} is a plot of the analytical CDF and PDF of $\Delta$ in Model I {with $\lambda=10^{-5}$, {$r_2=1/(60\sqrt{\lambda})$} and $\ell_0=1$}. {The simulation-based CDF is also plotted and closely matches the analytical as anticipated.} We observe that for the chosen parameters, the PDF of $\Delta$ in Model I is concentrated around $10^{-3} \leq \Delta \leq 1.6\times 10^{-3}$.}

\subsection{Model II: {Equidistant IRS Deployment}}

In this setup, we assume that {the tIRS is deployed equidistant from the tBS and tUE with} the distance $R_1=R_2=\frac{\sqrt{R_0}}{c}$. {For a given $R_0$ (and $R_1$)} there are two locations where the IRS can be placed in such a setup as shown in Fig. \ref{model2}. {From the cosine rule,} such a setup requires $c< \frac{2}{\sqrt{R_0}}$; we use {$c=\frac{2}{\sqrt{3 {\mathbb{E}[R_0]}}}$} where $\mathbb{E}[R_0]={(2\sqrt{q\lambda})}^{-1}$ {follows from \eqref{f_R0} \footnote{We require $R_0<\frac{4}{c^2}$; however, since our choice $c$ is based on $\mathbb{E}[R_0]$, not $R_0$, there is a probability of not meeting this criterion. With $c=\frac{2}{\sqrt{3 {\mathbb{E}[R_0]}}}$ this probability is less than $<0.1 \%$. Thus, our $c$ is acceptable and two locations for the IRS exist that allow $R_1=R_2=\frac{\sqrt{R_0}}{c}$.}.} With this $c$, $R_1= R_2=\frac{\sqrt{3 {\mathbb{E}[R_0]}{R_0}}}{2} $.

{Accordingly, the triangle parameter for Model II is $\Delta= c^{2\eta} \ell_0^{\eta}$. With the choice of $c$ above, $\Delta=\left(\frac{4 \ell_0}{3 \mathbb{E}[R_0]}\right)^{\eta}=\left(\frac{8\sqrt{q\lambda} \ell_0}{3}\right)^{\eta}$. Note that for Model II the triangle parameter is deterministic and only depends on a few constants.} The SIR can thus be written as \eqref{sinr_gen0} evaluated at  $\Delta=c^{2\eta} \ell_0^{\eta}$. {Note that as $\Delta$ is a constant in Model II, \eqref{c_k} is exact in this case. }



\section{Numerical Results}

In this section, unless mentioned otherwise, we consider BS intensity {$\lambda=10^{-5}$}, $\ell_0=1$ and $\eta=4$. {We select $\eta=4$ to represent some urban and suburban environments, $\ell_0=1$ is selected for simplicity in the path loss model \textcolor{magenta}{and $\lambda=10^{-5}$ is realistic for current dense cellular BS deployments}.} Each simulation consists of $10^5$ Monte Carlo runs. {The analytical no-IRS curves are obtained via \eqref{ccdfSIR_noIRS}.} Additionally, for Model I, unless mentioned otherwise, {$r_2=1/(60\sqrt{\lambda})$}.


\subsection{{Validation of Analysis}}

%
%
%


\begin{figure}
\begin{minipage}[htb]{\linewidth}
\centering\includegraphics[width=0.8\linewidth]{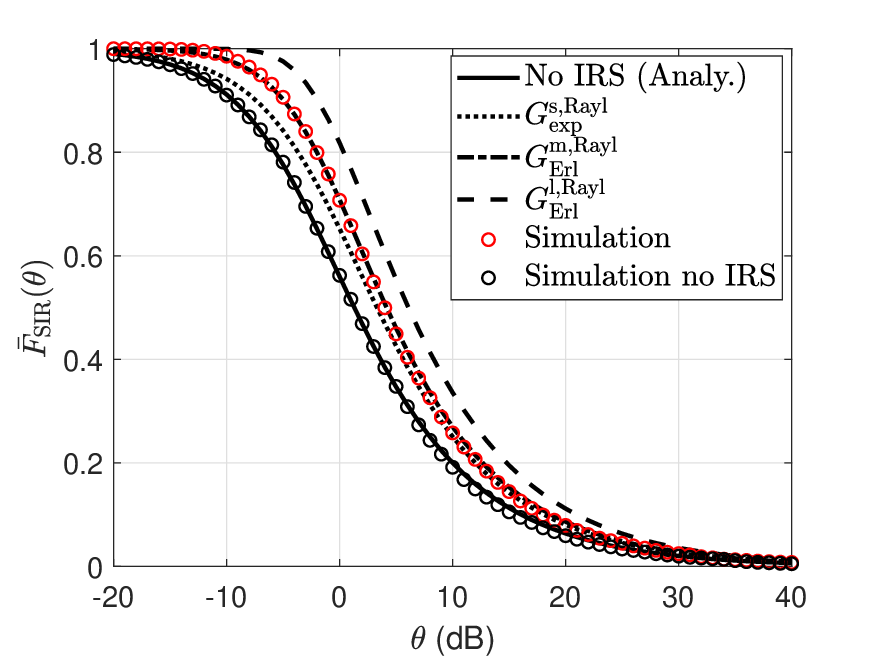}
\subcaption{$N=10$}\label{cvgM1_N10}
\end{minipage}\;
\begin{minipage}[htb]{\linewidth}
\centering\includegraphics[width=0.8\linewidth]{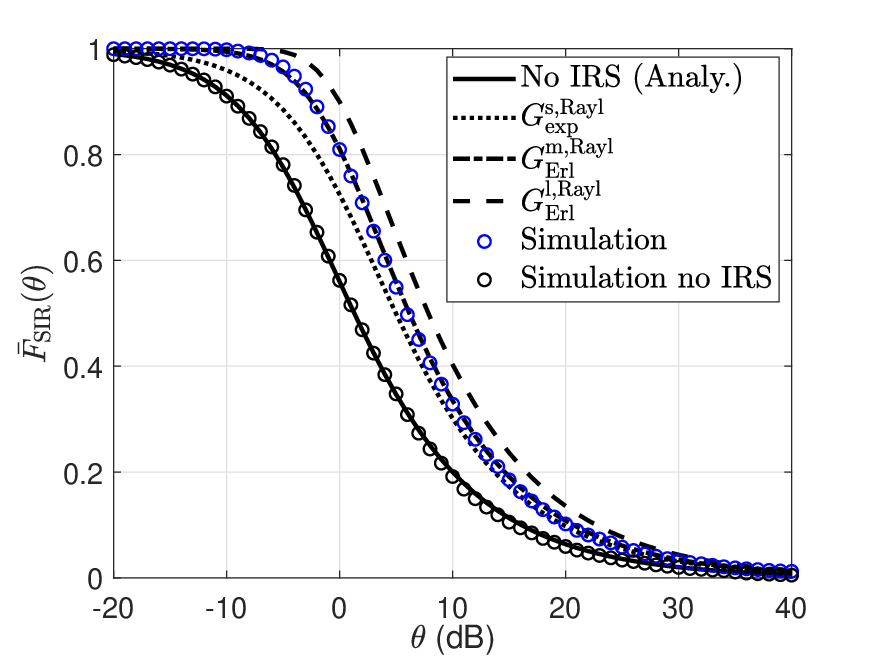}
\subcaption{$N=20$}\label{cvgM1_N20}
\end{minipage}\;
\begin{minipage}[htb]{\linewidth}
\centering\includegraphics[width=0.8\linewidth]{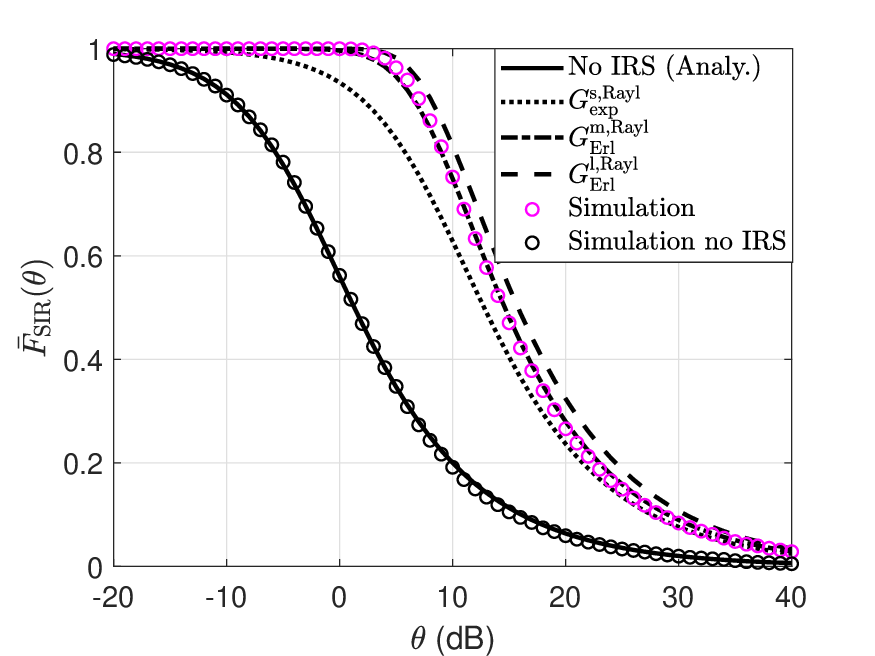}
\subcaption{$N=100$}\label{cvgM1_N100}
\end{minipage}
\caption{The SIR CCDF in Model I {with $\mu=1$} obtained via simulation and the analytical approximations.}\label{cvgM1}
\end{figure}

Fig. \ref{cvgM1} plots the SIR CCDF and throughput in Model I for different $N$ with {$\mu=1$ (i.e., Rayleigh fading on the D\&I links)}. The simulation results are compared with the analytical results for the SIR obtained using the CDF of {$G_{\rm exp}^{\rm s, Rayl}$, $G_{\rm Erl}^{\rm m, Rayl}$ and $G_{\rm Erl}^{\rm l, Rayl}$}. We also plot the analytical and simulation results for the no-IRS scenario. {As anticipated, the IRS-assisted cases always outperform communication without an IRS, and the performance improves with $N$. For the IRS-assisted scenario,} note from Fig. \ref{deltaStats} that for the chosen parameters, $\Delta$ is concentrated around $1.3 \times 10^{-3}$. The corresponding $N\Delta = 1.3 \times 10^{-2}$, $2.6 \times 10^{-2}$ and $1.3 \times 10^{-1}$ for $N=10$, $20$ and $100$, respectively. {From Table \ref{NDeltaTable},} the CDF of $G \to G_{\rm Erl}^{\rm m, Rayl}$ when $10^{-4} < N\Delta < 1$ and the CDF of $G \to G_{\rm Erl}^{\rm l, Rayl}$ when $N\Delta \geq 1$. In Figs. \ref{cvgM1_N10}, \ref{cvgM1_N20} and \ref{cvgM1_N100}, we observe that simulation-based SIR CCDF is closest to the curve using $G_{\rm Erl}^{\rm m, Rayl}$ for each $N$, validating Table \ref{NDeltaTable}. Further, as $N$ and therefore $N\Delta$ increases, the gap between the simulation and the curve using $G_{\rm Erl}^{\rm l, Rayl}$ decreases. {Using the CDF of $G_{\rm exp}^{\rm s, Rayl}$ results in a lower bound on the CCDF of the SIR in this regime of $\Delta$ and becomes looser as $N$ and therefore $N\Delta$ grows.} {Overall, using the CDF of $G_{\rm Erl}^{\rm m, Rayl}$ results in a very good match with the simulation for all the $N$ values making it a strong candidate for approximating the SIR CCDF for Model I with the chosen parameters.}


\begin{figure}
\begin{minipage}[htb]{\linewidth}
\centering\includegraphics[width=0.8\linewidth]{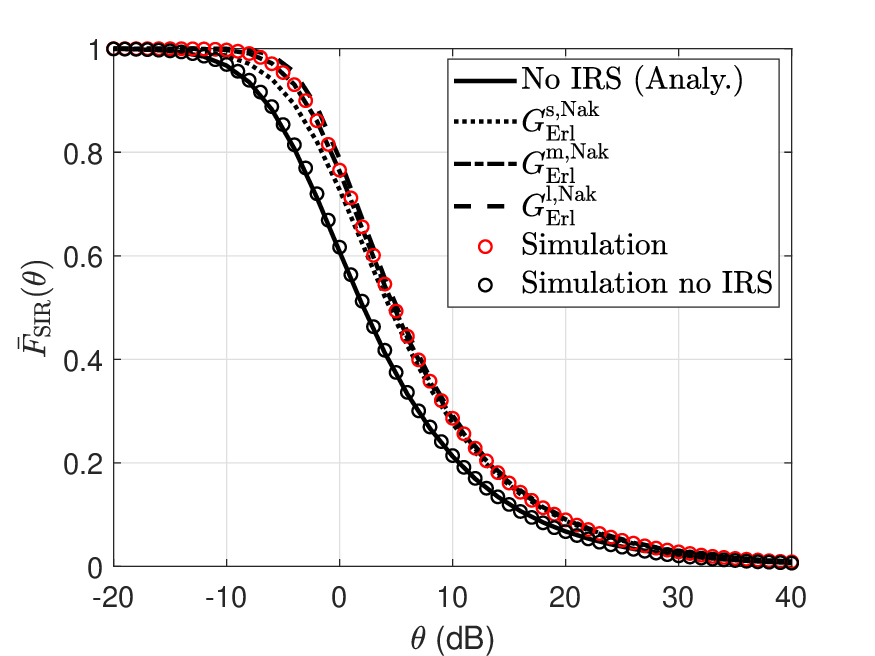}
\subcaption{$N=10$}\label{cvgM1_N10_NakMu2}
\end{minipage}\;
\begin{minipage}[htb]{\linewidth}
\centering\includegraphics[width=0.8\linewidth]{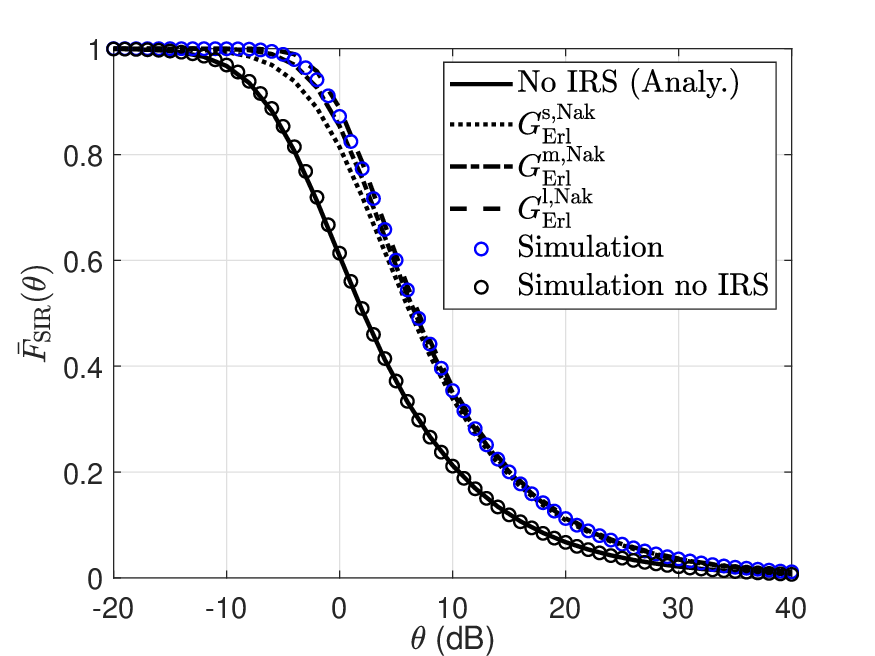}
\subcaption{$N=20$}\label{cvgM1_N20_NakMu2}
\end{minipage}\;
\begin{minipage}[htb]{\linewidth}
\centering\includegraphics[width=0.8\linewidth]{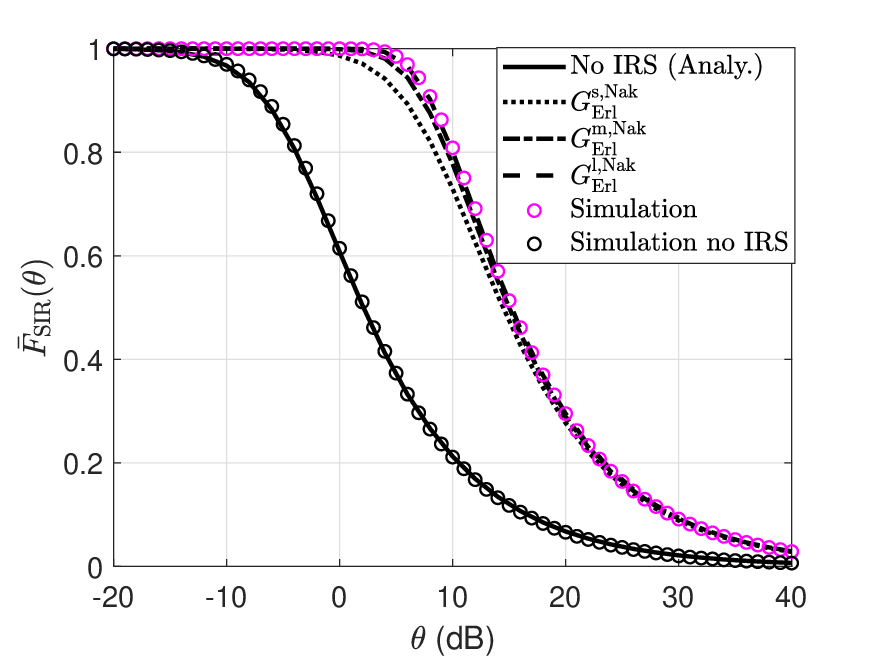}
\subcaption{$N=100$}\label{cvgM1_N100_NakMu2}
\end{minipage}
\caption{{The SIR CCDF in Model I {with $\mu=2$} obtained via simulation and the analytical approximations.}}\label{cvgM1_NakMu2}
\end{figure}

{ Fig. \ref{cvgM1_NakMu2}, like Fig. \ref{cvgM1}, plots the CCDF of the SIR in Model I for different $N$ for Nakagami fading with $\mu=2$ on the D\&I links. {Again, the performance improves with $N$}. Since the chosen parameters result in $\Delta$ concentrated around $1.3 \times 10^{-3}$, the corresponding $N\Delta$ all fall under $\chi=\rm m$ (cf. Table \ref{NDeltaTable}). {We also observe that} as $N$ increases, the SIR CCDF gets closer to $\chi= \rm l$. This is verified as the simulation-based SIR CCDF matches the analytical using $G_{\rm Erl}^{\rm m, Nak}$ for $N=10$ and $20$, while for $N=100$, it lies between $G_{\rm Erl}^{\rm m, Nak}$ and $G_{\rm Erl}^{\rm l, Nak}$ due to the larger $N\Delta$. In contrast to {$\mu=1$ (i.e., Rayleigh fading)} in Fig. \ref{cvgM1}, the analytical SIR CCDFs for {$\mu=2$} obtained via the three approximations of $G$ are much closer to each other. This highlights that channel hardening is also more {pronounced for higher $\mu$} and is corroborated by the lower variance {as $\mu$ grows} (cf. Fig. \ref{varY2}).  }

\begin{figure}
\begin{minipage}[htb]{\linewidth}
\centering\includegraphics[width=0.8\linewidth]{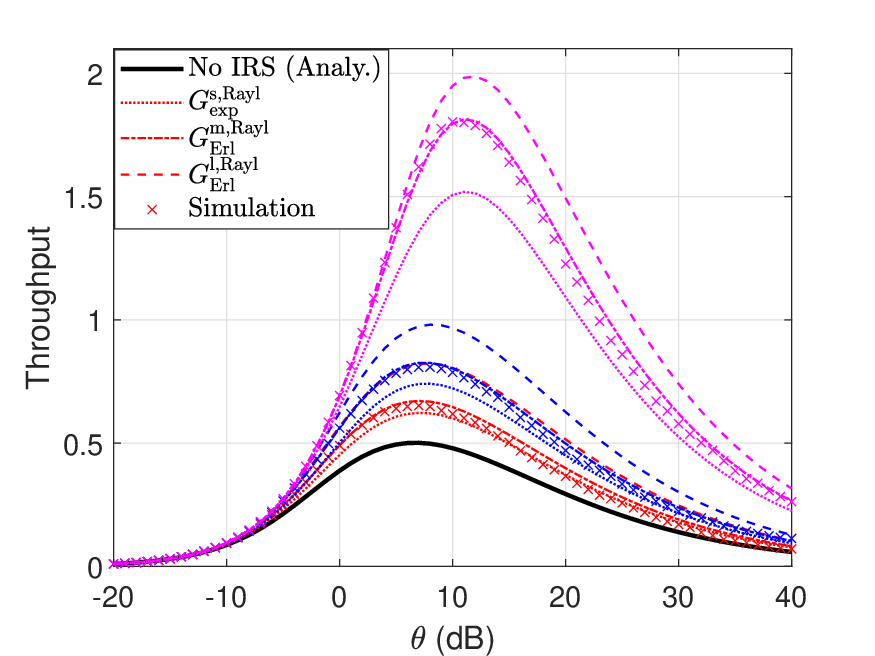}
\end{minipage}
\caption{The throughput in Model I {with $\mu=1$} obtained via simulation and the analytical approximations. Red (blue, magenta) represent $N=10$ ($20$, $100$).}\label{rateM1}
\end{figure}

{Fig. \ref{rateM1} is a plot of the throughputs for Model I {with $\mu=1$} corresponding to the SIR CCDF in Fig. \ref{cvgM1}. The simulation results obtained are compared to those obtained analytically using the CDF of {$G_{\rm exp}^{\rm s, Rayl}$, $G_{\rm Erl}^{\rm m, Rayl}$ and $G_{\rm Erl}^{\rm l, Rayl}$}. As $\Delta$ is concentrated around $1.3 \times 10^{-3}$, for these values of $N\Delta$, using the CDF of $G_{\rm exp}^{\rm s, Rayl}$ results in a lower bound, $G_{\rm Erl}^{\rm l, Rayl}$ in an upper bound and $G_{\rm Erl}^{\rm m, Rayl}$ {a good match for the simulation-based throughput.} The lower bound using $G_{\rm exp}^{\rm s, Rayl}$ becomes looser with $N$ while the upper bound using $G_{\rm Erl}^{\rm l, Rayl}$ becomes tighter with $N$. As anticipated, increasing $N$ improves the throughput for every value of $\theta$. Because of the trade off between reliability and throughput, each throughput curve has an optimum $\theta$. In particular, at the optimum $\theta$, IRS assistance with $N=10$ ($20, 100$) results in a {$31.6 \%$ ($63 \%, 263.7 \%$)} increase in throughput.}



\begin{figure}
\begin{minipage}[htb]{\linewidth}
\centering\includegraphics[width=0.95\linewidth]{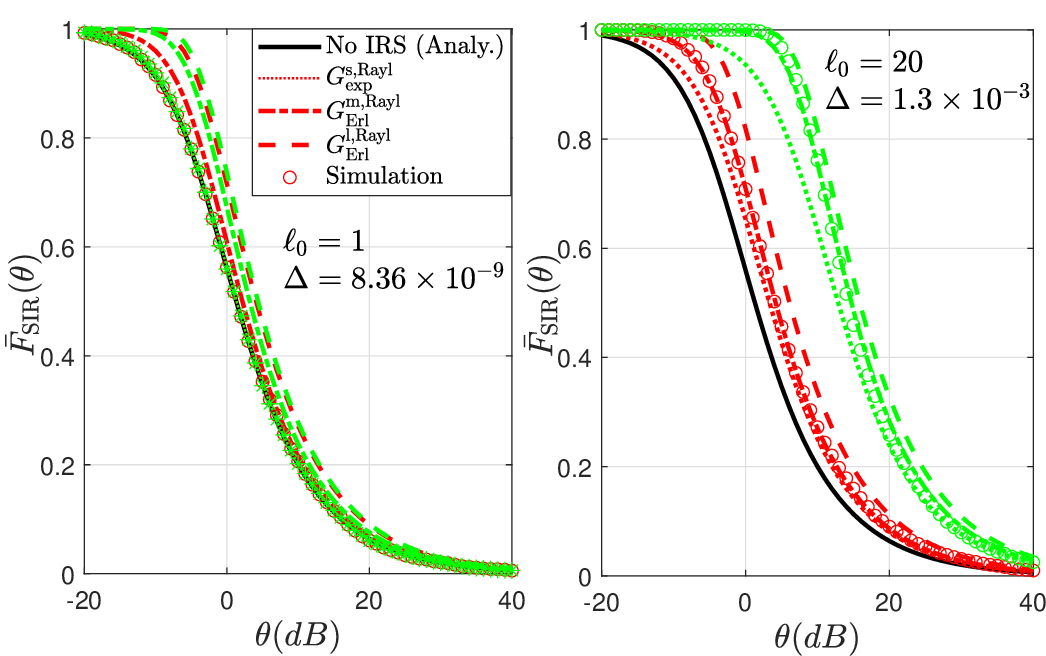}
\caption{{The SIR CCDF in Model II with $\mu=1$. Red (green) lines represent $N=10$ ($N=100$).}}\label{cvgM3both}
\end{minipage}
\end{figure}

{Fig. \ref{cvgM3both} plots the SIR CCDF for Model II with $\mu=1$ for $N=10$ and $100$. When $\ell_0=1$, $\Delta=8.36 \times 10^{-9}$; here IRS assistance does not improve performance from the case without IRS assistance and increasing $N$ too has a negligible impact on performance. {This occurs because $N\Delta<10^{-4}$ for both $N$ and the CDF of $G$ matches that of $G_{\rm exp}^{\rm s, Rayl}$ {(cf. Table \ref{NDeltaTable}) as in the no-IRS case}.} Further, due to the very low value of $\Delta$ in this scenario, no gains are seen from IRS assistance since the signal amplification term $\mathbb{E}[\tilde{G} \mid \Delta]$ {is very close to $1$}. This can be interpreted as an impact of the placement of the IRS and thus the corresponding large values of relative link distances $R_1/\ell_0$ and $R_2/\ell_0$ with the chosen parameters.} Due to the double path loss encountered in the indirect link coupled with the large values of the relative link distances, we see nearly no gains from IRS assistance here. 

{When $\ell_0$ is increased to $20$ in Fig. \ref{cvgM3both}, $\Delta=1.3 \times 10^{-3}$. In contrast to $\ell_0=1$, gains from IRS assistance are seen here. This is because while the distances remain large, the increase in $\ell_0$ decreases the relative distance to the BS and therefore improves path loss. {Additionally, while $\Delta$ in Model I is a RV, for the chosen parameters in Fig. \ref{cvgM1}, it was concentrated around $1.3 \times 10^{-3}$. {We} thus observe that the SIR CCDF in Fig. \ref{cvgM3both} is quite similar to that in Figs. \ref{cvgM1_N10} and \ref{cvgM1_N100}. This highlights the impact of $\Delta$ not just on performance, but as a measure of the model used for IRS placement.} As anticipated according to Table \ref{NDeltaTable}, the SIR CCDF obtained using $G_{\rm Erl}^{\rm m, Rayl}$ matches the simulation best for both $N$ for this $\Delta$. It should also be noted that if $\Delta$ had been increased by increasing $\lambda$ while keeping $\ell_0=1$, we would have the same improvement in SIR CCDF. This again highlights the significance of the triangle parameter $\Delta$ on performance.} 


\subsection{Impact of IRS Parameters}
\begin{figure}
\begin{minipage}[htb]{\linewidth}
\centering\includegraphics[width=0.95\linewidth]{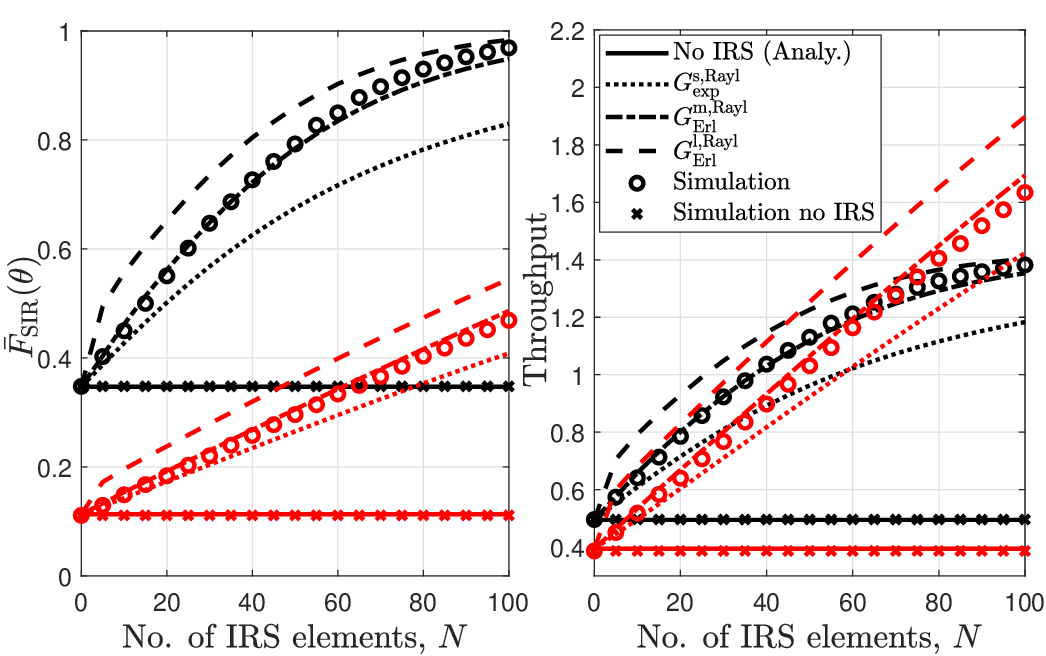}
\caption{{SIR CCDF and throughput for Model II using $\ell_0=20$, $\Delta=1.3 \times 10^{-3}$ {and $\mu=1$}. Black (red) lines represent $\theta=5$ dB ($15$ dB).}}\label{vsN}
\end{minipage}
\end{figure}

\begin{figure}
\begin{minipage}[htb]{\linewidth}
\centering\includegraphics[width=0.95\linewidth]{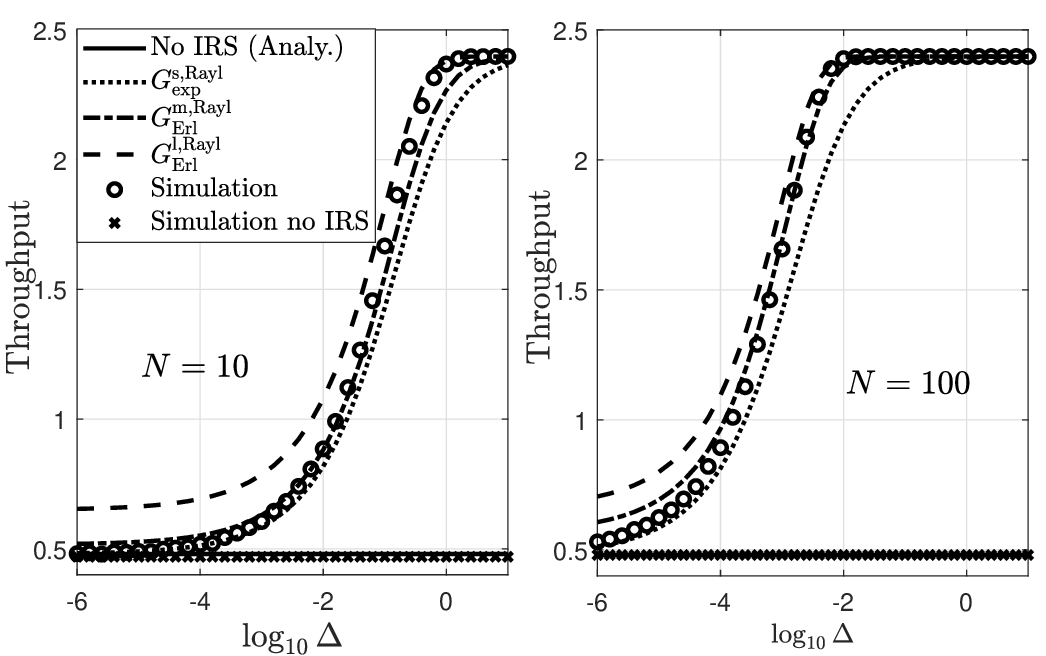}
\caption{{Throughput vs. $\Delta$ for Model II using $\theta=10$ dB {and $\mu=1$}.\\}}\label{rateVsDelta}
\end{minipage}
\end{figure}

Fig. \ref{vsN} is a plot of the SIR CCDF and throughput for Model II {with $\mu=1$} as $N$ increases using $\theta=5$ and $15$ dB. Both the simulation and analytical results using the CDF of $G_{\rm exp}^{\rm s, Rayl}$, $G_{\rm Erl}^{\rm m, Rayl}$ and $G_{\rm Erl}^{\rm l, Rayl}$ are plotted. {With the chosen parameters, $\Delta=1.3 \times 10^{-3}$ thus, with growing $N$, and therefore $N\Delta$, we see the simulation curve for the SIR CCDF go from being closer to $G_{\rm Erl}^{\rm m, Rayl}$ to moving towards $G_{\rm Erl}^{\rm l, Rayl}$. Further, the gap between the simulation and $G_{\rm exp}^{\rm s, Rayl}$ curve grows with $N$ and therefore $N\Delta$.} 
{The throughput for each curve deploying IRS grows with $N$. Additionally, the $\theta=5$ dB case outperforms the $\theta=15$ dB case at low $N$, as its relatively higher SIR CCDF dominates the throughput in the low $N$ regime.} However, at higher $N$, the SIR CCDF of $\theta=15$ dB becomes large enough so that its higher transmission rate results in larger throughput, and consequently it outperforms $\theta=5$ dB. Interestingly, for $\theta=5$ dB, {after first growing linearly with $N$,} the throughput starts saturating at higher $N$. {On the other hand,} for $\theta=15$ dB, in the plotted regime of $N$, the growth in throughput is linear with $N$. {In general,} the throughput grows \emph{at most} linearly with $N$ {even when the optimum $\theta$ is selected}. {The throughput growth being proportional to $N$ and not $N^2$ is in agreement with the fact that the BS radiates power in all directions and that the IRS and UE only capture a fraction of this power, as well as the findings in \cite{IRS0}.}


Fig. \ref{rateVsDelta} is a plot of the simulation-based and analytical throughput using the CDFs of $G_{\rm exp}^{\rm s, Rayl}$, $G_{\rm Erl}^{\rm m, Rayl}$ and $G_{\rm Erl}^{\rm l, Rayl}$ {for $\mu=1$}. We observe that the throughput grows monotonically with $\Delta$ and saturates when the {SIR CCDF} approaches 1 as fixed rate transmissions are being used. {Note, however, that due to the superiority of the system with the larger $N$, the throughput for $N=100$ saturates to the maximum at a lower $\Delta$ than for $N=10$}. Thus, we explicitly observe that increasing each of $N$ and $\Delta$ improves the performance of an IRS-aided network. {The simulation-based throughput goes from matching the curve for $G_{\rm exp}^{\rm s, Rayl}$ to $G_{\rm Erl}^{\rm m, Rayl}$ and then $G_{\rm Erl}^{\rm l, Rayl}$ as $\Delta$ and therefore $N\Delta$ grows along the lines of Table \ref{NDeltaTable}. Note that the match with $G_{\rm exp}^{\rm s, Rayl}$ at low $\Delta$ is observed more easily for $N=100$ while the match with $G_{\rm Erl}^{\rm l, Rayl}$ is observed more easily for $N=10$ due to the gap between simulations and $G_{\rm exp}^{\rm s, Rayl}$ ($G_{\rm Erl}^{\rm l, Rayl}$) becoming looser (tighter) with $N$ as seen in Fig. \ref{rateM1}.}


{While Fig. \ref{rateVsDelta} was plotted for Model II, Model I (or any other model)} with the same $\Delta$ values would result in the same performance. In Model II, $\Delta$ can be varied by varying $\lambda$ (and consequently $R_0$) or $\ell_0$. On the other hand, in Model I, $\Delta$ can also be varied by changing the direct link distances such as the fixed distance $R_2$. This highlights that the triangle parameter $\Delta$ is a measure of important metrics such as IRS placement, network density and relative path loss that impact the performance of an IRS-aided network.

\begin{figure}[h]
\begin{minipage}[htb]{\linewidth}
\centering\includegraphics[width=0.8\linewidth]{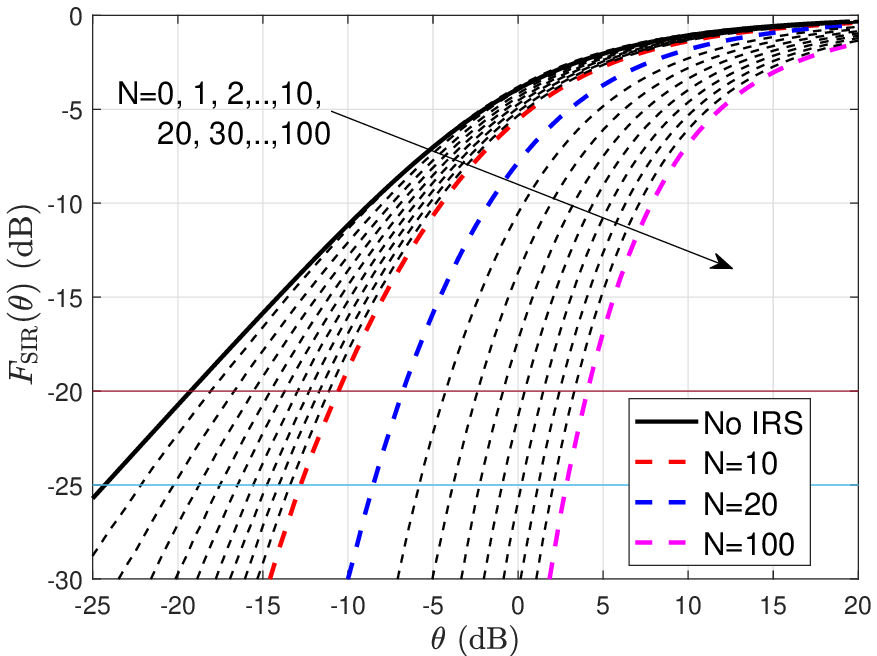}
\caption{SIR CDF (simulation) in dB for Model I with $r_2=5$, {$\mu=1$.}}\label{logOut}
\end{minipage}
\begin{minipage}[htb]{\linewidth}
\centering\includegraphics[width=0.8\linewidth]{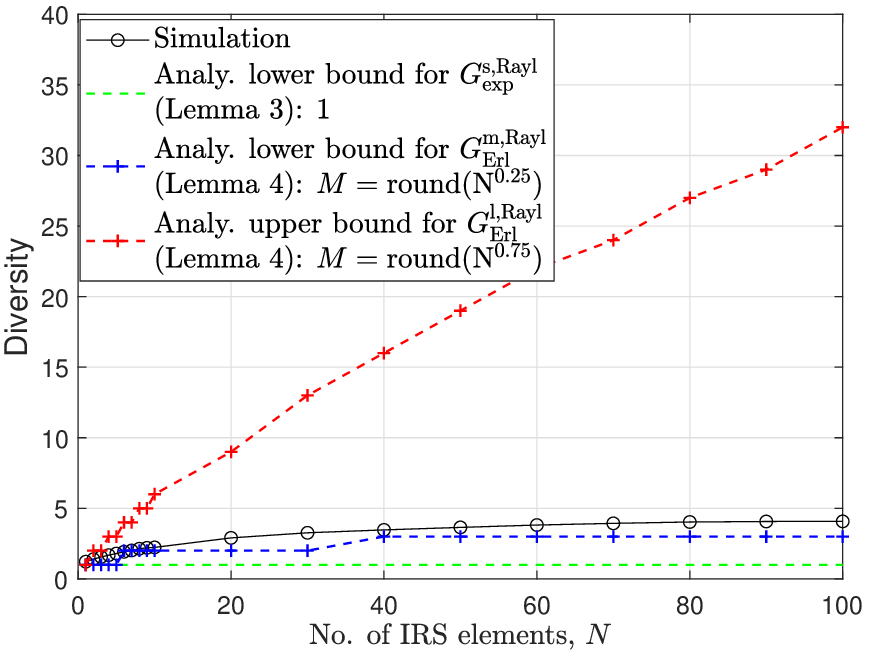}
\caption{Diversity vs. $N$ for Model I with $r_2=5$ {and $\mu=1$.}}\label{diversityVsN} 
\end{minipage}
\end{figure}

{Fig. \ref{logOut} is a plot of the simulation-based SIR CDF in dB  vs. the SIR threshold $\theta$ in dB {with $\mu=1$} for Model I with $r_2=5$ for a wide range of $N$ from $0$ to $100$. We observe that as $N$ increases, the slope of each curve as $\theta \to 0$, {which corresponds to the diversity gain \eqref{diversityDef},} increases. Fig. \ref{diversityVsN} plots the diversity based on the simulation results in Fig. \ref{logOut}; the slope is measured between $F_{\rm{SIR}}(\theta)=-20$ dB and $-25$ dB for each $N$. As $\Delta$ is concentrated around $1.3 \times 10^{-3}$ for the chosen parameters, the CDF of $G_{\rm Erl}^{\rm m, Rayl}$ is a good approximation for all the plotted $N$ in this scenario. {We observe that the analytical bound on the diversity calculated in Lemma 4 for $G_{\rm Erl}^{\rm m, Rayl}$, $M=\text{round}(N^{0.25})$, is a lower bound on the diversity in the case of Rayleigh fading on the D\&I links {(i.e., $\mu=1$)}. Naturally, the lower bound on the diversity from Lemma 3, i.e., $d \geq 1$ also holds, but it is looser. We also plot the bound on the diversity from Lemma 4 for $G_{\rm Erl}^{\rm l, Rayl}$, $M=\text{round}(N^{0.75})$, and find that it is an upper bound on the diversity in this scenario. The lower bound from $G_{\rm Erl}^{\rm m, Rayl}$ is much tighter than the upper bound from $G_{\rm Erl}^{\rm l, Rayl}$ because the lower bound is derived from the $G_{\rm Erl}^{\rm m, Rayl}$ approximation, which is closer to $G$ in this scenario.} {Deploying IRSs can substantially improve performance; with the parameters in Fig. \ref{diversityVsN} we observe that a diversity gain of at least {3.9} is attainable when $N=100$.} {Further, we note that the additional gain in diversity from increasing $N$ is diminishing continuously. {For instance, increasing $N$ from 10 to 20 results in a 30.23 \% increase in diversity, while increasing it from 90 to 100 only results in a mere 0.163 \% increase.} Hence, further increasing $N$ when $N$ is already large will not improve fading conditions as much as it would when $N$ is smaller.} 
 
 \begin{figure}
\begin{minipage}[htb]{\linewidth}
\centering\includegraphics[width=0.8\linewidth]{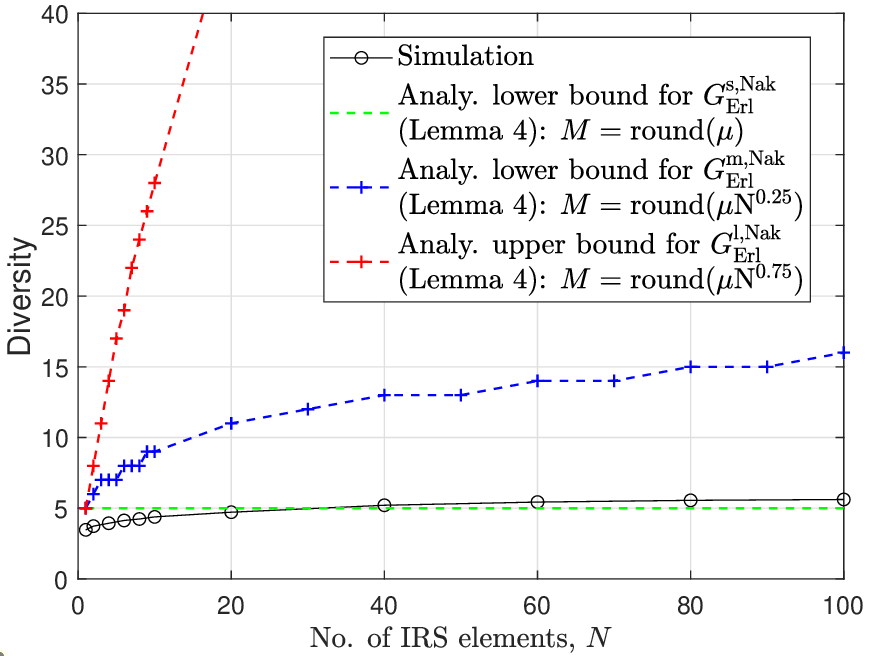}
\caption{Diversity vs. $N$ for Model I with $r_2=5$ {and $\mu=5$}.}\label{diversityVsN_Nak}
\end{minipage}
\end{figure}

\begin{figure}
\begin{minipage}[htb]{\linewidth}
\centering\includegraphics[width=0.8\linewidth]{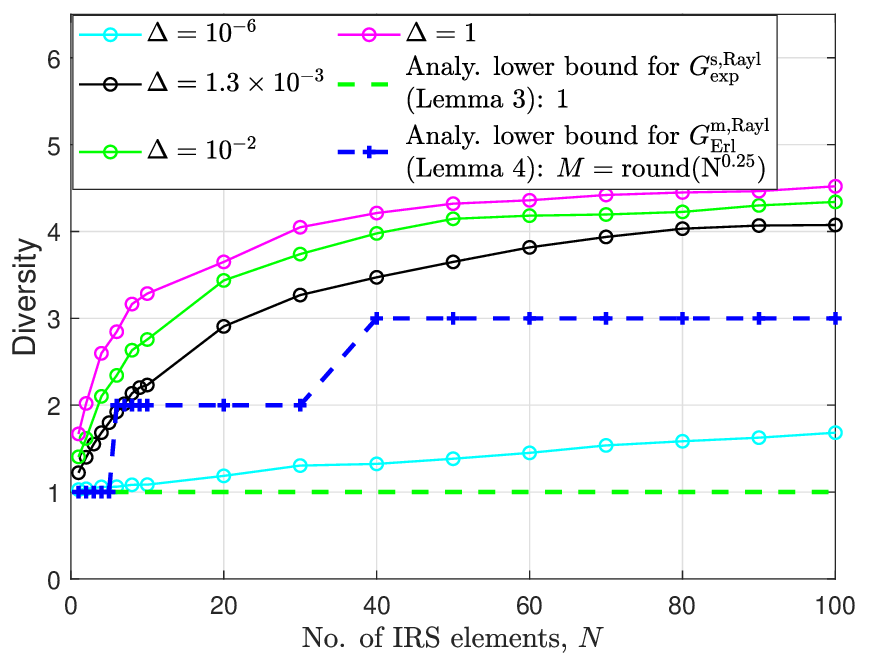}
\caption{Diversity vs. $N$ in Model II with {$\mu=1$}.}\label{divVsN_diffDelta} 
\end{minipage}
\end{figure}
 
 Fig. \ref{diversityVsN_Nak} is a plot of the simulation-based diversity vs. $N$ {with $\mu=5$} for Model I with $r_2=5$. The diversity increases with $N$ as expected. Since $\Delta$ is concentrated around $1.3 \times 10^{-3}$ for the chosen parameters, the CDF of $G_{\rm Erl}^{\rm m, Nak}$ is a good approximation for all the plotted $N$ in this scenario. {This} corresponds to the diversity bound $M=\text{round}(\mu N^{0.25})$ from Lemma 4. We observe that unlike the case of {$\mu=1$ (Rayleigh fading)}, $\text{round}(\mu N^{0.25})$ is an upper bound in this case. We also plot the diversity bounds from Lemma 4 corresponding to $G_{\rm Erl}^{\rm s, Nak}$ and $G_{\rm Erl}^{\rm l, Nak}$ which are $M=\text{round}(\mu)$ and $M=\text{round}(\mu N^{0.75})$, respectively. As anticipated, $\text{round}(\mu N^{0.75})$ is a looser upper bound. Interestingly, in the case of {$\mu>1$}, even in the small $N\Delta$ regime which uses $G_{\rm Erl}^{\rm s, Nak}$, we have an upper bound on the diversity. This is because $\text{round}(\mu)$ is an upper bound in the case of small $N$ even for the plotted $N\Delta$ values (which are in the mid range of $N\Delta$). {Thus,} for the smaller $N\Delta$, $\text{round}(\mu)$ is an upper bound. {As in the case of {$\mu=1$}, the additional gain in diversity from increasing $N$ diminishes continuously.}
 


{Fig. \ref{divVsN_diffDelta} plots the simulation-based diversity vs. $N$ {with $\mu=1$} using different $\Delta$ in Model II. We also show the {analytical bound on the diversity for $G_{\rm Erl}^{\rm m, Rayl}$, $M=\text{round}(N^{0.25})$, from Lemma 4 and the lower bound from Lemma 3. As anticipated, diversity increases with $\Delta$ as the fading conditions get better. When $N\Delta$ is large enough so that $G$ is approximated by $G_{\rm Erl}^{\rm m, Rayl}$ or $G_{\rm Erl}^{\rm l, Rayl}$, the lower bound for $G_{\rm Erl}^{\rm m, Rayl}$, $M=\text{round}(N^{0.25})$, holds. However, when $\Delta$ is very low, such as in the case of $\Delta=10^{-6}$ for all the plotted $N$, $G$ is worse than $G_{\rm Erl}^{\rm m, Rayl}$, and $\text{round}(N^{0.25})$ is not a lower bound on the diversity. When $\Delta$ is as low as $10^{-6}$, the CDF of $G$ matches that of $G_{\rm exp}^{\rm s, Rayl}$ for all the plotted $N$. The analytical lower bound on the diversity when using $G_{\rm exp}^{\rm s, Rayl}$ is $1$. Thus, despite the very low $\Delta$ value, we still observe that the diversity for the $\Delta=10^{-6}$ curve slowly increases with $N$. As the diversity increases with $N\Delta$, the lower bound from Lemma 3 holds for all $N\Delta$ values.} {We also find that for a given $N$, the additional gain in diversity from increasing  $\Delta$ is diminishing continuously. This shows that increasing $\Delta$ when $\Delta$ is larger will not improve fading conditions as much as it would when $\Delta$ is smaller.} 


\section{Conclusion}

In this paper a large IRS-assisted cellular network is studied. {The geometry of such a network is characterized by the triangle parameter $\Delta$, which is a function of the distance in the BS-IRS-UE triangle and the path loss exponent. The effect of the indirect link via the IRS can be captured using the ECPC $G$ and a signal amplification factor. We find that the proposed approximation for the CDF of $G$ depends only on $N\Delta$. These proposed statistics are also used to obtain bounds on the diversity gain. We show that in addition to $N$, $\Delta$ plays an important role on IRS performance; {increasing $\Delta$ (by adjusting the geometry of the BS, IRS and UE) increases the signal amplification factor.} This highlights the impact on performance of factors like IRS placement, network density, relative link distances and path loss in the indirect link, which $\Delta$ captures.}

{We show that performance is independent of the IRS placement model used if $\Delta$ is the same for a given $N$. In fact, performance is solely impacted by $N$ and $\Delta$; if both of these are very small, we see no gains from IRS deployment. However, for larger values of $N$ and $\Delta$ {(such as $N\Delta>10^{-4}$)}, significant gains can be attained from IRS deployment: with {$\mu=1$} and $\Delta=1.3 \times 10^{-3}$ for $N=10$ ($20, 100$), {up to} a $31.6 \%$ ($63 \%$, $263.7 \%$) increase in throughput from the no-IRS case {was observed}. We also find that the additional gain in diversity from both increasing $N$ and $\Delta$ diminishes continuously. {The insights obtained from our results provide guidelines on how to judiciously select the network parameters and geometry.} 

{Our model and the triangle parameter lend themselves to some interesting and relevant extensions. For instance, incorporating the impact of non-isotropic BSs, which result in more or less power in the direction of a node such as the IRS or UE, can be captured as a change in link distance and therefore geometry. Similarly, the impact of the IRS reflection not being in the exact direction, which also leads to a power loss, can be captured as a change in $R_2$ and therefore $\Delta$.}} {Another important extension of this work is to study the distribution of the ECPC $G$ in the presence of multiple antennas at the BS and at the UE. In addition to the signal amplification factor, such a setup would also result in beamforming gains, improving performance further.} {Lastly, it will be interesting to explore the impact of correlated fading between the elements of the IRS.} The signal amplification and distributions of $G$ would be impacted by such correlation. It would be valuable to see how much correlation impacts IRS-aid and to investigate if the size of the IRS elements can be adjusted to make this impact negligible.

\appendices

\section{Proof of Lemma 1}\label{L1proof}
By expanding the terms of $\tilde{G}$ defined in \eqref{Y} we have
{\small \begin{multline*}
\mathbb{E}[\tilde{G} \mid \Delta]= \mathbb{E}\left[ g_0^2  \right] + \mathbb{E}\left[ \sum_{i=1}^N \Delta g_{i,1}^2 g_{i,2}^2 \right] +  2 \mathbb{E}\left[  \sum_{i=1}^N g_0 \Delta^{\frac{1}{2}}  g_{i,1} g_{i,2} \right] \\ +  2\mathbb{E}\left[ \sum_{i=1}^N \sum_{j=i+1}^N \Delta g_{i,1} g_{i,2} g_{j,1} g_{j,2}   \right]\\
 \stackrel{(a)}= \mathbb{E}\left[ h_0 \right]  +  \sum_{i=1}^N \Delta \mathbb{E}\left[h_{i,1} \right] \! \mathbb{E}\left[h_{i,2} \right] + 2 \!  \sum_{i=1}^N \! g_0 \Delta^{\frac{1}{2}}  \mathbb{E}\left[g_{i,1} \right] \! \mathbb{E}\left[ g_{i,2} \right] \\+ 2 \! \sum_{i=1}^N \! \sum_{j=i+1}^N \! \Delta  \mathbb{E}\left[ g_{i,1} \right] \! \mathbb{E}\left[ g_{i,2} \right] \! \mathbb{E}\left[g_{j,1} \right]\! \mathbb{E}\left[ g_{j,2}   \right] \\
 \stackrel{(b)}= \mathbb{E}\left[  h_0 \right] + 2  N \Delta^{\frac{1}{2}} \mathbb{E}\left[  g_0 \right]  \mathbb{E}\left[g_{i,1} \right] \mathbb{E}\left[g_{i,2} \right]  + N  \Delta \mathbb{E}\left[  h_{i,1} \right] \mathbb{E}\left[  h_{i,2} \right]  \\ + 2  \frac{N(N-1)}{2} \Delta  \mathbb{E}\left[ g_{i,1} \right]^2 \mathbb{E}\left[ g_{i,2} \right]^2 . 
\end{multline*} }
As $g_0$, $g_{i,1}$, $g_{i,2}$ are i.i.d. {Nakagami RVs with parameters $\mu$ and $\omega=1$}, the first two terms of (a) are obtained using their respective squares $h_0$, $h_{i,1}$ and $h_{i,2}$ which are i.i.d. unit-mean {gamma} RVs {with parameters $\mu$ and $1/\mu$}. Due to the independence of $g_{i,1}$ and $g_{i,2}$ $\forall i$, we obtain the last two terms in (a) and because of their identical distribution, we obtain the last two terms in (b). We obtain \eqref{E_Y} from the distributions of these RVs. \qed}

\section{Proof of Lemma 3}\label{L3proof}
By plugging \eqref{cdfSIR_gen} into \eqref{diversityDef} we have 

{\small
\begin{align*}
&d \!=\! \lim_{\substack{\theta \to 0}}\! \frac{\log \! \mathbb{E} \! \left[\! \left(\! 1 \!-\! {\exp \left(\!  - \theta \mu_1 \gamma_1 I      \!\right)} \! \right) \! \right] }{\log  \theta} \! \stackrel{(a)}\geq \! \lim_{\substack{\theta \to 0}} \! \frac{\mathbb{E} \! \left[\!  \log \!   \left( \! 1 \!-\! {\exp \left(\!  - \!\theta \mu_1 \gamma_1 I   \!\right)} \! \right) \!\right] }{\log \theta} \\
& \stackrel{(b)}= \lim_{\substack{\theta \to 0}}   \mathbb{E} \left[  \frac{  \mu_1 \gamma_1 I  \exp \left(  - \theta \mu_1 \gamma_1 I  \right)   }{\theta^{-1} \left( 1 - \exp \left(  - \theta \mu_1 \gamma_1 I  \right)  \right) } \right] \\
&\!\stackrel{(c)}= \! \lim_{\substack{\theta \to 0}}  \!  \mathbb{E} \! \left[\! \gamma_2 \frac{  \mu_1 \gamma_1 I   e^{  - \theta \mu_1 \gamma_1 I      \!} \!-\! (\!\mu_1 \gamma_1 I \!)^2 \theta e^{ - \theta \mu_1 \gamma_1 I    }    }{  \mu_1 \gamma_1 I  e^{ - \theta \mu_1 \gamma_1 I  } } \! \right]  \! \stackrel{(d)}=\!   \mathbb{E} \left[ \! \frac{  \mu_1 \gamma_1 I }{\mu_1 \gamma_1 I} \! \right] \! =\! 1.
\end{align*}}
Here $(a)$ follows from Jensen's inequality {for a concave function}. We obtain $(b)$ and $(c)$ by applying L'Hopital's rule. By setting $\theta \to 0$, $(d)$ follows and we obtain Lemma 3. \qed 

\section{Proof of Lemma 4}\label{L4proof}
{By plugging \eqref{cdfSIR_genErl} into \eqref{diversityDef} we have 
{\small \begin{multline*}
d= \lim_{\substack{\theta \to 0}} \frac{\log \mathbb{E} \left[ 1 - \sum\limits_{k=0}^{M-1} \frac{1}{k!} e^{- \theta  \gamma_E I}   \left( \theta   \gamma_E I \right)^k \right] }{\log \theta} \\
 {\stackrel{(a)}\gtreqless} \lim_{\substack{\theta \to 0}} \frac{\mathbb{E} \left[ \log  \left( 1 - e^{- \theta  \gamma_E I}  - \sum\limits_{k=1}^{M-1} \frac{1}{k!} e^{- \theta  \gamma_E I}  \left( \theta   \gamma_E I \right)^k \right) \right] }{\log \theta} \\
\stackrel{(b)}= \lim_{\substack{\theta \to 0}}  \! \mathbb{E} \! \frac{ e^{- \theta  \gamma_E I} \! \left(\! \gamma_E I   \!-\!  \sum\limits_{k=1}^{M-1} \! \frac{\left(\gamma_E I \right)^k \theta^{k-1}}{(k-1)!}       \!  - \! \frac{\left(\gamma_E I \right)^{k+1} \theta^k}{k!}    \!\right) }{\theta^{-1} \! \left(\! 1 \!-\! e^{- \theta  \gamma_E I}  \!-\! \sum\limits_{k=1}^{M-1} \frac{1}{k!} e^{- \theta  \gamma_E I}  \left( \theta   \gamma_E I \right)^k \! \right)   } \\
 \stackrel{(c)}= \lim_{\substack{\theta \to 0}}  \! \mathbb{E} \!  \frac{  \frac{\left(\gamma_E I \right)^{M} \theta^{M}}{(M-1)!}           }{ {e^{ \theta  \gamma_E I}}  \!- \! 1  \! - \!  \sum\limits_{k=1}^{M-1} \frac{\left( \theta   \gamma_E I \right)^k}{k!}  } \\
 \! \stackrel{(d)}= \! \lim_{\substack{\theta \to 0}}   \mathbb{E}\! \frac{M \left(\!\gamma_E I \!\right)^M \theta }{  \left(\! \gamma_E I \!\right)^{M\!-\!1} \! e^{\theta \gamma_E I} \! - \! \left(\!\gamma_E I \!\right)^{M\!-\!1} } \! \stackrel{(e)}=\! \lim_{\substack{\theta \to 0}} \!  \mathbb{E}\! \frac{M \left(\!\gamma_E I \!\right)^M  }{  \left(\! \gamma_E I \!\right)^M \! e^{\theta \gamma_E I}  } \!  \stackrel{(f)}=\! M.
\end{multline*} }
Here {$(a)$ is an upper or lower bound according to Jensen's inequality (cf. Remark 5)}. By applying L'Hopital's rule $(b)$ is obtained and rewritten as $(c)$. L'Hopital's rule is applied $(M-1)$ times to obtain $(d)$, and applied once again to obtain $(e)$. By setting $\theta \to 0$, $(f)$ is obtained. As $M$ is independent of $R_0$, $R_1$, $R_{\rm d}$ and $I$, the expectation is dropped in $(f)$. \qed}


\section{Proof of Proposition 1}\label{P1proof}

From Fig. \ref{model1} and \eqref{R1}, $|R_0-r_2| \leq R_1 \leq R_0+r_2$. Let us define $\alpha=R_0^2+ r_2^2$ and $\beta=2R_0 r_2$. Then, we have 
\[C=R_1^2=\alpha- \beta \cos \phi. \]
Using the notation $D= \beta \cos \phi$, $D$ has the same distribution as Jake's spectrum which is used in wireless communication to determine the random frequency shift of a time varying channel \cite{jakeSpectrum_Book}. Accordingly, the distribution of $D$ is
\begin{align*}
F_{D}(x) &= {\pi}^{-1} \sin^{-1} \left( {\beta^{-1} x} \right) + 0.5 , \; -\beta \leq x \leq \beta 
.
\end{align*}
As $C=\alpha-D$, $F_{D}(x)=1-F_{C}(\alpha-x)$ and the CDF of $C$ is
\begin{align*}
F_C(x) = 0.5 - {\pi^{-1}} \sin^{-1} \left( {\beta^{-1} (\alpha-x)} \right), \;\;\; \alpha-\beta \leq x \leq \alpha+\beta.
\end{align*}
Since $C=R_1^2$, $F_{C}(x)=F_{R_1}(\sqrt{x})$. Substituting in the values of $\alpha$ and $\beta$, we obtain the CDF in \eqref{F_R1}. \qed} 

\bibliographystyle{IEEEtran}
\bibliography{References}

\end{document}